\documentclass[prd, 11pt, a4paper, showpacs, floatfix, groupedaddress, notitlepage, dvipsnames, reprint, twosided, onecolumn, preprintnumbers]{revtex4-1}

\usepackage{geometry}
\usepackage[utf8x]{inputenc}
\usepackage[activate={true,nocompatibility},final,tracking=true,kerning=true,factor=1500,stretch=10,shrink=10]{microtype}

\usepackage[colorinlistoftodos]{todonotes}
\usepackage{graphicx}
\usepackage{bm, amssymb, amsmath, amsfonts}
\usepackage{multirow}

\usepackage{hyperref}
\hypersetup{
colorlinks = true, 	    
linkcolor = magenta,	
citecolor = blue,		
}
\usepackage[linesnumbered,ruled,lined,boxed]{algorithm2e}
\usepackage[noend]{algpseudocode}

\SetAlCapSty{xAlCapSty}

\newcommand{\atcp}[1]{\tcp*[r]{\parbox[t]{.575\linewidth}{#1\hfill}}}

\SetCommentSty{mycommfont}

\SetNlSty{mynlfont}{}{} 

\RestyleAlgo{algoruled}

\usepackage{caption}
\usepackage{subcaption}

\def \Hz	    {\mathrm{Hz}}

\def \avgSNRLoss {\langle {\delta \rho}/\rho \rangle}
\def \GW        {\texttt{GW}}
\def \LIGO      {\texttt{LIGO}}

\def \SNR       {\text{SNR}}
\def \GstLAL    {\texttt{GstLAL}}
\def \LLOID     {\texttt{LLOID}}
\def \CBC       {\texttt{CBC}}
\def \BNS       {\texttt{BNS}}
\def \BBH       {\texttt{BBH}}
\def \NSBH      {\texttt{NSBH}}
\def \IMBH      {\texttt{IMBH}}
\def \SVD       {\texttt{SVD}}
\def \QR        {\texttt{QR}}

\def \RP        {\texttt{RP}}
\def \RMF       {\texttt{RMF}}
\def \MP        {\texttt{MP}}
\def \MB        {\texttt{MB}}
\def \GB        {\texttt{GB}}
\def \RAM       {\texttt{RAM}}
\def \GRB       {\texttt{GRB}}
\def \GBM       {\texttt{GBM}}
\def \EM        {\texttt{EM}}
\def \KAGRA     {\texttt{KAGRA}}
\def \Virgo     {\texttt{Virgo}}
\def \min       {\textrm{min}}
\def \sinc      {\textit{sinc}}
\def \rank      {\textrm{rank}}
\def \th        {\texttt{th}}
\def \NP        {\texttt{NP}}
\def \stage     {\texttt{Stage}}
\def \TaylorT   {\texttt{TaylorT4}}
\def \HTC       {\texttt{HTC}}
\def \SPARK     {\texttt{SPARK}}
\def \Johnson   {\texttt{Johnson}}
\def \Lindenstrauss {\texttt{Lindenstrauss}}

\def \Error     {\texttt{Error}}
\def \residual  {\texttt{residual}}
\def \streamer  {\texttt{streamer}}
\def \Apache    {\texttt{Apache}}
\def \Split     {\texttt{split}}
\def \bank      {\texttt{bank}}
\def \Stage     {\texttt{Stage}}
\def \Gstreamer {\texttt{Gstreamer}}
\def \HTC       {\texttt{HTC}}
\def \HPC       {\texttt{HPC}}
\def \O         {\texttt{O}}
\def \l         {\texttt{L}_{2}}
\def \LSC {\texttt{LSC}}
\def \JL {\texttt{JL}}

\setcitestyle{numbers,square}

\begin{document}

\preprint{LIGO DCC number \color{blue}{\bf LIGO-P2100004}}

\title{
\mbox{
Random projections in gravitational-wave searches of compact binaries II:} \\ efficient reconstruction of the detection statistic
}

\author{Amit~Reza}			\email[]{amit.reza@iitgn.ac.in}

\author{Anirban~Dasgupta}	\email[]{anirbandg@iitgn.ac.in}

\author{Anand~S.~Sengupta}	\email[]{asengupta@iitgn.ac.in}

\affiliation{
\mbox{Indian Institute of Technology Gandhinagar, Gujarat 382355, India.}
}


\begin{abstract}
Low-latency gravitational wave search pipelines such as $\GstLAL$ take advantage of low-rank factorization of the template 
matrix via singular value decomposition \mbox{(\SVD)}. The matrix factors can be used to 'reconstruct' the detection statistic
to the desired precision by linearly combining the vectors obtained from filtering the data against the top-few basis vectors
that span the template vector space. With unprecedented improvements in detector bandwidth and sensitivity in advanced-$\LIGO$
and Virgo detectors, one expects the size of template banks to increase by several orders of magnitude in upcoming searches.
Naturally, this poses a formidable computational challenge in factorizing extremely large template matrices. Previously,
[Kulkarni~{\em{et~al.}} \cite{kulkarni2019random}], we had introduced the idea of random projection-based matrix factorization
as a computationally viable alternative to \mbox{\SVD}, applicable for large template banks. In this follow-up paper,
we demonstrate the application of a block-wise randomized matrix factorization ($\RMF$) algorithm for computing low-rank 
factorizations at a preset average fractional loss of $\SNR$. This new scheme is shown to be more efficient in the context of 
the $\LLOID$ framework of the $\GstLAL$ search pipeline. Further, it is well-known that for very large template banks, 
the total computational cost of the search is dominated by the cost of reconstructing the detection statistic as compared to 
that of filtering the data. However, the issue of optimizing the reconstruction cost has not been addressed satisfactorily so
far in the available literature. We show that it is possible to approximately reconstruct the time-series of the matched-filter
detection statistic at a fraction of the total cost using the matching pursuit algorithm. The combination of the two algorithms
presented in this paper can handle online searches involving large template banks more efficiently. We have analyzed the total
computational cost in detail and offer various tips for optimally applying the $\RMF$ scheme in different parts of the parameter
space. The algorithms presented in this paper are designed in a suitable manner that can be efficiently implemented over a
distributed computing architecture. Results from several numerical simulations have been presented to demonstrate their 
efficacy.

\vspace{12pt}
\textbf{Keywords} - Gravitational-wave (\GW), \GstLAL, Singular Value Decomposition (\SVD), Random Projection (\RP), Randomized matrix factorization (\RMF), Computational Complexity

\end{abstract}

\pacs{}
\maketitle 

\section{Introduction}
\label{Sec:Intro}

%

The detection of a short gamma-ray burst ($\GRB$-$170817$A) \cite{abbott2017gravitational} by the Gamma-ray Burst Monitor ($\GBM$) instrument aboard the Fermi satellite, coincident with the discovery of the $\GW$-$170817$~\cite{abbott2017gw170817} event in data from the advanced-$\LIGO$ and Virgo detectors event marks an important epoch in $\GW$-astronomy: that of an era of multi-messenger observations.
More multi-messenger discoveries can provide a better understanding of sources of $\GW$s, and the underlying astrophysical processes leading to their progeny. 
It is obvious that the prompt detection of the $\GW$s is very crucial to chase the transient events associated with the $\GW$s sources. Hence the design of efficient and real-time detection pipelines has become essential.  
%


While some recently proposed machine-learning based detection pipelines have been demonstrated to be quite efficient at near real-time detection of astrophysical $\GW$ signals, most of the well-established, traditional 
data analysis pipelines are based on the matched-filtering ~\cite{helstrom1994elements, Satya} scheme. 
%
Under the matched-filtering analysis framework, one computes the cross-correlation between a large number of theoretically modelled $\CBC$ waveforms (called the template bank) and the detector output, which may contain astrophysical $\GW$ signals from coalescing compact binaries buried in instrumental and environment noise that couple to the apparatus. The cross-correlation is computed over the sensitive bandwidth of the detector and weighted inversely by the noise power spectrum of the detector. For additive Gaussian noise, the matched filter technique can be shown to be optimum in that it yields the maximum signal-to-noise ratio ($\SNR$). While the noise in real $\LIGO$ or $\Virgo$ detectors is neither wide-sense stationary nor Gaussian, a phenomenological final detection statistic using the matched-filtering output is employed to ensure near optimally.

New $\GW$ detectors such as $\LIGO$-India, $\KAGRA$ are slated to join the global network of terrestrial $\GW$ detectors in the next few years. The combined data from such a network can not only help reconstruct the physical properties of the $\CBC$ sources with unprecedented accuracy but also help test new and subtle phenomena. It is reasonable to expect that a larger number of next-generation $\GW$ detectors will bring in new challenges to the matched-filtering based detection pipelines, especially in their ability at low-latency detection of transient signals: (a) the most straightforward challenge will arise from the large volume of data that will be needed to be analyzed in real-time (b) improvement in detector sensitivity at lower frequencies, and the deployment of theoretical template waveforms that include precession effects will increase the number of templates in the bank by several orders of magnitude, (c) the improvement of detector sensitivity at low frequencies will lead to many more cycles of $\GW$ signals within the detector's sensitive bandwidth, which in turn will increase the time duration of these transient signals. For example, the signal from a typical coalescing binary neutron-star system will last several tens of minutes, which will have a precipitous effect on the amount of data samples that need to be analyzed. The issues outlined above set the stage for the analysis presented in this paper, where we explore alternative, more economical implementations of the matched filtering scheme useful for the real-time $\GW$ searches from $\CBC$ sources.

The development of a low-latency search pipeline ($\GstLAL$) ~\cite{cannon2012toward, messick2017analysis, sachdev2019gstlal} by the $\LSC$ has significantly reduced the computational cost of implementing a matched filtering search leading to a reduction in latency of detection. This has also facilitated rapid $\EM$ follow-up observations. The $\GstLAL$ pipeline calculates the approximate matched-filtering $\SNR$  based on the truncated $\SVD$ \cite{golub1981numerical} framework~\cite{cannon2010}. The basic idea is as follows: instead of using the templates directly for filtering the data, a set of basis vectors spanning the template bank is pre-computed, and only a fraction of the basis vectors (in descending order of their corresponding singular values) are used to filter the data. In other words, the top-few basis vectors are used as surrogate templates to filter the data, which are then combined to approximately reconstruct the matched-filter $\SNR$ time-series. Such an ad-hoc truncation of the bases inevitably leads to a loss in $\SNR$. Operationally, one works with enough number of bases to guarantee an average $\SNR$ loss $\avgSNRLoss \simeq 10^{-3}$. The power of this method comes from the rapid decay in the spectrum of singular values, which enables us to use only a very small fraction of bases vectors to reach the desired accuracy in $\SNR$, thereby leading to a significant reduction in the filtering cost.

The time complexity of $\SVD$ factorization does not scale well with the increase in matrix size leading to several practical problems in its implementation for $\GW$ searches with very large template banks. It is a formidable computational challenge to $\SVD$ factorize a very large template bank in its entirety. As a practical workaround, such large template banks are split into sub-banks and the $\SVD$-based matched filtering scheme is applied independently to each of the split banks~\cite{messick2017analysis, sachdev2019gstlal}. However, such a strategy diminishes the linear dependence of the template waveforms, and as a consequence, the anticipated computational advantage of the $\SVD$ method is substantially compromised. In a recent work \cite{kulkarni2019random}, we have shown that the computation of the basis vectors for an entire template bank is possible by implementing a probabilistic low-rank matrix approximation framework which can adequately address the scaling issue of $\SVD$ factorization of huge template matrices. We have shown theoretically that the random projection ($\RP$) \cite{dasgupta1999} based matrix factorization technique can be a practical alternative for computing the set of basis in the $\GstLAL$ search pipeline.
Randomized matrix factorization ($\RMF$) algorithms ~\cite{halko2011, martinsson2016} 
are computationally more efficient, numerically stable, highly parallelizable, and can be fine-tuned to work optimally for the problem identified above. It incurs fewer floating-point operations than the standard deterministic matrix factorization as it uses a fixed amount of passes over the data.

The average fractional loss of $\SNR$ under the truncated $\SVD$ method is directly proportional to the discarded number of singular values. Depending on the target average fractional loss of $\SNR$, one can decide the required number of basis vectors. Hence, a $\SVD$ based filtering scheme allows using an acceptable average fractional loss of $\SNR$ as a tunable parameter to determine the search efficiency using discarded set of basis vectors based on the less weighted singular values. 

However, the truncated $\SVD$ technique can measure the average fractional loss of $\SNR$ corresponding to the number of discarded basis vectors. Hence, the whole spectrum of singular values and basis vectors must be used to fix the cut-off on the average fractional loss of $\SNR$.  For a large template bank, where the number of required basis vectors is significantly less in comparison to the total number of templates corresponding to a fixed average fractional $\SNR$ loss, the number of singular values and basis vectors that need to be discarded are is the large fraction of the number of templates. Hence computation and storing of the whole set of singular values and basis vectors is a waste of computational resources, which is a computationally expensive step and slows down the entire analysis, and needs to be optimized to make the whole process faster. It is more effective to reverse the problem where one can find a fixed number of basis vectors based on predefined average fractional $\SNR$ threshold, and as a result, there is no need to obtain the less important singular and basis vectors. But using the truncated $\SVD$ framework, it is hard to address this issue for a large template bank. Hence, in this work, we introduce block-wise $\RP$ based matrix factorization of a template matrix that can easily handle this kind of issue and is able to find an optimal number of bases based on a predefined average fractional $\SNR$ loss. The basic idea of block-wise $\RMF$ is to obtain a few essential singular values and basis vectors in each iteration depending on the size of the block of the $\RP$ matrix and check for the optimal criteria on the average fractional $\SNR$ loss. Once it reaches the predefined fractional $\SNR$ loss, there is no need to project further. Mathematically, the block-wise projection helps obtain a fixed importance sub-space of the template matrix in each iteration. The basis vectors can be computed from that sub-space only. If the block size is small, one has to use a small sub-space to calculate the basis vectors in each iteration.


The $\SVD$ based matched filtering scheme's primary focus has been to reduce the filtering cost. The set of basis vectors are firstly computed, and filtering is performed against those bases. Since the number of bases is significantly less than the number of templates, the computational cost is saved. But, this is not the complete scenario as here we are discarding the cost of reconstruction of the $\SNR$ time-series. After getting the filter output from the correlation between the data and basis, we need to multiply it with the coefficient vectors to obtain the $\SNR$ time-series for each template. This reconstruction is an issue for a large bank as the cost of reconstructing the coefficient is high. Hence, just reducing the filtering cost does not suffice as the ultimate solution to this problem. We need to handle the issue of the reduction of the reconstruction cost too, which is challenging as well as an open research problem. To the best of our knowledge, there are no research articles available to address this issue. In this work, we are providing an approach to reduce the reconstruction cost for the first time.
The method obtains a sparse coefficient corresponding to a specific template waveform based on the matching pursuit ($\MP$) algorithm. This sparse coefficient can be used to reconstruct the $\SNR$ time-series for each of the templates. The $\MP$ algorithm transforms the original dense coefficient vectors into sparse ones. Therefore, these transforms can reduce the reconstruction cost as the number of floating-point operations will be less than the use of the dense coefficient matrix. Thus, a randomized matrix factorization scheme and matching pursuit algorithms together can increase the power of this kind of basis-based filtering scheme, and in a distributed set-up, it can provide an optimal framework compared to the $\SVD$ based filtering scheme.   

The rest of the paper is organized as follows. In section-\ref{Sec:CBC}, we have summarized the matched filtering scheme for the $\CBC$ searches, the $\SVD$ based matched filtering scheme, and the total computational cost analysis of this scheme. In section-\ref{Sec:RMF}, the conceptual idea behind the $\RMF$ has been described. The $\RMF$ method based on a predefined average $\SNR$ loss has been described in sub-section-\ref{Subsec:RMF with Error} which is followed by a section ( section-\ref{Sec:MP_Algo}) in which reconstruction of the \SNR time-series using $\MP$ has been proposed. The optimal $\RMF$ schemes for $\GstLAL$ pipeline have been designed in section-\ref{Sec:LLOID-RMF}. In the following sections, details of the advanced $\RMF$ method and a new approach for reducing the reconstruction cost combining $\RMF$ with $\MP$ have been described.
\section{Compact Binaries searches}
\label{Sec:CBC}
The coalescence of compact binaries \text{e.g.} neutron stars, black holes, is a promising source of $\GW$ signal. The standard matched filtering technique is used as an initial step to identify a signal's presence in a detector output. The matched filtering operation has to be performed between the detector output and theoretical waveform. A large number of theoretically well-modelled waveforms are needed to probe the component mass parameter space for this matched filtering operation. This set of theoretical waveforms are called the {\em template waveforms} $\bm{h}^{\alpha}(t): \alpha = 1,2,\cdots N_{T}$ and can be thought of as vectors lying in a vector space. The number of time-series samples determines the dimension of each vector $(N_{s} = f_{s} \times T)$, where $f_{s}$ and $T$ are the sampling frequency and time, respectively. Suppose the number of template waveforms for a fixed parameter space is $N_{T}$. 
Each template $\bm{h}^{\alpha}(t)$ consists of two orthogonal template $\bm{h}^{\alpha}_{0}$ and $\bm{h}^{\alpha}_{\pi/2}$ such that 
\begin{equation}
\bm{h}^{\alpha}(t) = \mathcal{A}_{0} \, \bm{h}^{\alpha}_{0} \big( t, \theta \big) + \mathcal{A}_{\pi/2} \, \bm{h}^{\alpha}_{\pi/2} \big( t, \theta \big) \, , 
\label{Eq:Waveform-decom}
\end{equation}
where $\theta$ represents the set of intrinsic and extrinsic parameters, hence a parameter space consisting of $N_{T}$ number of templates can be considered as a template matrix $\mathbf{H}$ of size $2N_{T} \times N_{s}$, where normalized whitened template waveforms are stacked row-wise. The template waveforms are whitened by the power spectral density of the noise. Therefore, we can define the matched-filter output for a $\alpha^{\th}$ template at a specific time against detector output $\bm{s}$ as follows:
\begin{equation}
\rho_{\alpha}(\Delta t) = \sqrt{\big( H_{2\alpha -1} \, \cdot \bm{s}^{T} \big)^{2} \, + \, \big( H_{2 \alpha} \, \cdot \bm{s}^{T} \big)^{2} } \, ,
\label{Eq:MatchedFilter}
\end{equation}
where, $H_{\alpha}$ denotes the $\alpha^{\th}$ row of template matrix $\mathbf{H}$ and $\bm{s}^{T}$ implies the transpose of the detector output $\bm{s}$. Note that the Eq.\ref{Eq:MatchedFilter} shows the matched filter output for a fixed time-stamp. It is always preferable to compute the correlation between the whitened template and detector output in a frequency domain wherein the computational cost of matched filtering with a specific template against data considering all time-stamps is $N_{s} \, \log N_{s}$. Whereas, to obtain the matched filter output for all time-stamps in a time domain, the template waveform has to be shifted by $\Delta t$ every time to obtain the output defined in Eq.\ref{Eq:MatchedFilter}. This shifted version of each of the rows of $\mathbf{H}$ can be done by constructing a circulant matrix structure for each of the row vectors. Hence, in the time domain, the matched filter output for a fixed template with all the time stamps can be redefined as follows: 
\begin{equation}
\rho_{\alpha}(\Delta t) = \sqrt{\big(\mathcal{C}( H_{2\alpha -1}) \, \cdot \bm{s}^{T} \big)^{2} \, + \,  \big(\mathcal{C}(H_{2 \alpha}) \, \cdot \bm{s}^{T} \big)^{2} } \,,
\label{Eq:TDMatchedFilter}
\end{equation}
where $\mathcal{C}$ represents a circulant matrix for a specific row of $\mathbf{H}$. The dimension of $\mathcal{C}$ is $N_{s} \times N_{s}$. Hence, the computational cost for matched filter output for all time-stamps defined in Eq.\ref{Eq:TDMatchedFilter} is ${N_{s}}^{2}$. But this cost can be reduced to $N_{s} \, \log N_{s}$ as circulant matrix can be factored into discrete Fourier components. Hence, the computational cost to obtain the matched output for all time-stamps using frequency domain or time domain representation is the same, and it requires $N_{s} \, \log N_{s}$ floating-point operations. So, the total computational time of matched filtering scheme considering all the templates is $N_{T} \, N_{s} \, \log N_{s}$. In such a search, for a large template matrix $\mathbf{H}$, for example, one having the number of rows (templates) and the number of columns (time samples) of $\mathcal{O}(10^{5})$,  $\mathcal{O}(10^{6})$ respectively, the matched filtering task becomes very expensive as the number of floating-point operations will be of order $10^{12}$. 
For the upcoming searches, it is expected that the number of template waveforms and the number of time samples will increase significantly, \text{e.g.}, it can be of order $\mathcal{O}(10^{5})$, $\mathcal{O}(10^{6})$ respectively. Therefore the computational cost of performing matched filter operation will also be increased, which eventually makes the search process slow and tedious. The $\SVD$ based matched filtering scheme \cite{cannon2010} can reduce the computational cost of matched filtering. In the $\SVD$ based matched filtering scheme, a set of essential basis vectors have been computed first from a group of template waveforms. Further, data (\text{i.e.}, detector output) is filtered against these important bases only. Since the number of essential bases is less than the total number of template waveforms, thus it can reduce the computational cost of performing matched filters significantly. The basis representation of the template waveforms and matched filter operations in terms of basis and data is shown briefly in the next section. 



\subsection{Singular Value decomposition based Matched Filtering}
\label{Subsec:SVD-MF}
$\SVD$ can be used to compute the matched filtering output more efficiently. Instead of calculating the cross-correlation between the data vector $\bm{s}$ and template vector $\bm{h}^{\alpha}$ directly, the idea is to compute a set of orthogonal basis from template vectors using $\SVD$ and perform cross-correlation of the data stream with those set of basis vectors. 

Using $\SVD$, one can numerically compute a set of orthogonal basis ($\bm{v}_{\nu}$) in such a way that a specific linear combination of these basis vectors can represent the template vectors. After applying $\SVD$, the template matrix $\mathbf {H}_{2N_{T} \times N_{s}}$ can be decomposed into three special matrices $\mathbf{U}$, $\mathbf{\Sigma}$, and $\mathbf{V}$, where $\mathbf{V}$ and $\mathbf{U}$ consist of a set of orthonormal vectors and $\mathbf{\Sigma}$, the diagonal matrix contains the singular values in descending order of magnitude. $\mathbf{V}$ is a matrix of orthonormal bases whose each column are basis vectors. 
\begin{equation}
H_{\mu j} = \sum_{\nu = 1}^{2N_T}{u_{\mu \nu}\sigma_{\nu \nu} \, {v_{\nu j}}^{T}} = \sum_{\nu = 1}^{2N_T}{A_{\alpha \nu}\bm{v}_{\nu}} \, ,
\label{Eq:SVD decom}
\end{equation}
where $ \bm{v}_{\nu} = \{\bm{v}_{1},\bm{v}_{2},...,\bm{v}_{2N_T}\}$ is a set of basis vector. $A_{\alpha \nu}$ is the reconstruction matrix and each row defines the corresponding weights of the basis vectors for representing the row vectors of template matrix $\mathbf{H}$.
From Eq.(\ref{Eq:SVD decom}), it is clear that the required number of basis vectors for representing any row vector of the template matrix $\mathbf{H}$ is equal to the full-rank of the template matrix \text{i.e.} $2N_{T} = \min \big(2N_{T}, N_{s}\big), \, \text{if} \, 2N_{T} < N_{s}$.

For the non-precession and aligned spin waveform, the neighbouring template vectors are differed by phase only. Hence all of them are almost similar. Therefore, only a few (let us suppose $\ell$, where $\ell \ll 2N_{T}$) number of basis vectors are sufficient to reconstruct the template vectors, and indeed it will approximate the matched filter output with high precision. These $\ell$ number of basis vectors are the top most important singular directions, ranked based on the top-$\ell$ singular values. Hence it is possible to approximately reconstruct the template matrix $\mathbf{H}'$ based on this set of bases. The reconstructed matrix $\mathbf{H}'$ can be written as follows:
\begin{equation}
H'_{\mu j} = \sum_{\nu = 1}^{\ell}{A_{\alpha \nu}\bm{v}_{\nu}}
\label{Eq:TSVD}
\end{equation}
Exclusion of $(2N_{T} - \ell)$ number of less important basis vectors culminates the approximation in the length of each template vector. Hence, the overall effect will be observed on the preservation of the matrix norm (energy) of $\mathbf{H}$, which indeed turns out to find out the optimal value of $\ell$ for which the approximated template matrix ($\mathbf{H}'$) captures the maximum energy, \text{i.e.} $\| \mathbf{H} - \mathbf{H}'\|_{F} < \epsilon$. The spectra of the singular values (diagonal elements of the matrix $\mathbf{\Sigma}$) can help dictate the right choice of $\ell$. If singular values fall sharply, then it is enough to take $\ell \ll \min \big(2N_{T}, N_{s}\big)$, and based on the cut-off on the spectra, one can compute the value of $\ell$. This reduction of the number of basis vectors directly affects the matched filtering output $\rho(\Delta t)$. 
Thus the reconstruction of $\rho( \Delta t)$ based on topmost singular values can be defined as follows:
\begin{equation}
\rho'_{\alpha}(\Delta t) = \sqrt{\sum_{\nu = 1}^{\ell}{A_{(2\alpha-1)\nu}\big(\bm{v}_{\nu} \cdot \bm{s}^{T}\big)} + \sum_{\nu = 1}^{\ell}{A_{2\alpha \nu}(\bm{v}_{\nu} \cdot \bm{s}^{T})}}
\label{Eq:TSNR}
\end{equation}
The expected average fractional $\SNR$ loss ($\avgSNRLoss$) can be defined as a function of truncated singular values as defined in \cite{cannon2010}. 
\begin{equation}
\avgSNRLoss = \frac{1}{4N_{T}}\, \Bigg(\frac{\sum_{\nu = \ell + 1}^{2N_{T}}{\sigma_{\nu}^{2}}}{\sum_{\nu = 1}^{2N_{T}}{\sigma_{\nu}^{2}}}\Bigg)
\label{Eq:AvgSNRloss}
\end{equation}
Therefore average expected fractional loss of $\SNR$ could be used as a tuning parameter to fix a certain threshold for deciding the number of reduced basis vectors to reconstruct the original template waveforms. 
From Eq.(\ref{Eq:TSNR}), it is clear that we can approximate the $\SNR$ time-series calculation corresponding to each template waveform using only $\ell$ number of filtering operations between the top-$\ell$ basis vectors and the detector output. The time complexity of the filtering is $\mathcal{O}(\ell \, N_{s} \, \log N_{s})$, is very less in comparison to the required time complexity of the direct correlation between template vectors and data vectors because of $\ell \ll \min \big( 2N_{T}, N_{s} \big)$. In this way, reducing the set of basis vectors helps to reduce the total number of the matched filtering operations and, indeed, the time complexity of filtering. Note that this computational cost excluded the cost of reconstructing the $\SNR$ time-series after performing the filtering operation of the bases against data. From Eq.\ref{Eq:TSNR}, it is clear that to obtain every $\SNR$ time-series corresponding to each template waveform; one has to multiply with the corresponding coefficient vectors $A_{2\alpha-1}$ and $A_{2\alpha}$. For a small template matrix $\mathbf{H}$, the reconstruction cost is negligible, but for a large template matrix, it is not suggested to ignore the computational cost of reconstruction of the $\SNR$ time-series. Hence, we have demonstrated the complete cost analysis in the next section, including reconstruction cost and matrix factorization cost for the $\SVD$ based matched filtering scheme.

\subsection{Computational cost analysis of \SVD-based matched filtering approach}
\label{Subsec:Cost_SVD_Filter}
As mentioned above, the total computational cost of matched filtering using $\SVD$ can be decoupled as a sum up of two separate costs. 
\begin{enumerate}
\item {\color{blue}{Filtering cost}} is associated with calculating the cross-correlation values of the set of basis vectors with the data vectors. The total cost amounts to $\ell \, N_{s} \, \log N_{s}$.
\item {\color{blue}{Reconstruction cost}} is associated with the reconstruction of $\SNR$ time-series for each template. This involves multiplying each filter output with the corresponding coefficient vector. 
The total reconstruction cost is $2N_{T} \,\ell \, N_{s}$, as the coefficient matrix $A_{\alpha \nu} \in \mathbb{R}^{2N_{T} \times \ell}$, and the filter output matrix $\{\mathbf{V} \cdot \bm{s}^{T}\} \in \mathbb{R}^{\ell \times N_{s}}$. 
\end{enumerate}
Apart from the above two costs, there is another cost for performing matrix factorization of the template matrix $\mathbf{H}$ using $\SVD$. This computational cost is $\mathcal{O}(N_{s} \, {N_{T}}^{2}) \, : \, 2N_{T} < N_{s}$. Note that matrix factorization cost can be thought of as $\texttt{off}$-$\texttt{line}$ cost, which is a part of data pre-processing, whereas the filtering and reconstruction cost is an $\texttt{online}$ cost. But, it is crucial that this cost is included because for a large template matrix, if the factorization cost is huge, then computation of the basis vectors beforehand becomes an impossible task. For example, for a template matrix  $\mathbf{H}$ of size $10^{5} \times 10^{6}$, factorization using $\SVD$ is also unbearable as the time complexity will increase drastically as $\mathcal{O}(10^{16})$. Also, applying $\SVD$ to such a huge matrix requires a large run-time memory space which is also difficult to obtain. Therefore, it is clearly an issue because of the required time complexity and large memory space.

It is clear from the above discussion that the $\SVD$ based matched filtering approach adds an extra computational burden in terms of the reconstruction cost of the $\SNR$ time-series, whereas the direct matched filtering between template and data has only the filtering cost. So, comparing only the filtering cost for both methods is inequitable. The comparison of filtering cost using these two approaches is only reasonable if the reconstruction cost in case of $\SVD$ based matched filtering scheme is negligible, which is only possible if the number of templates in a template bank is few, which is an improbable scenario for the upcoming $\CBC$ searches. Therefore, it is crucial to investigate the possible ways to reduce the reconstruction cost. 

If the template matrix contains a large number of template waveforms, then the size of the coefficient matrix is large; therefore, the reconstruction cost of the $\SNR$ will be very high compared to the filtering cost, and hence reconstruction cost will dominate over the filtering cost. For example, consider size of $\mathbf{H}$ as $10^{5} \times 10^{6}$. Now if $\ell = 10^{4}$, then the filtering cost becomes $\mathcal{O}(10^{10})$, whereas the reconstruction cost of \SNR \ time-series becomes $\mathcal{O}(10^{15})$. Hence, the reconstruction cost is $10^{5}$ times the filtering cost. In fact, for any large template matrix, the reconstruction cost is always dominant no matter how many basis vectors are considered. In that scenario, only reducing the filtering cost will not be sufficient. Hence, it is a constraint on the matrix factorization-based matched filtering scheme. Currently, no method is available to reduce this reconstruction cost. In this work, for the first time, we prescribe an efficient solution procedure to address this issue, as is presented in detail in section \ref{Sec:MP_Algo}.

Additionally, to calculate the $\avgSNRLoss$ after considering a fixed number ($\ell$) of importance basis vectors for the approximation of \SNR \ time-series, firstly, it is required to compute all the basis vectors ($2N_{T}$) and their corresponding singular values. It is then required to fix a threshold for $\avgSNRLoss$ and choose those many singular values for which the threshold can be achieved. Only those sets of basis are considered as an important basis to compute the \SNR \ time-series. In practice, generally  $\ell \ll 2N_{T}$. Rest of $2N_{T} - \ell$ bases are considered as less important, and hence it is discarded from the final set of important bases. As these sets of bases are not used for further computation of \SNR \ time-series calculation, computation of those bases is wastage of computational resources and wastage of large memory space required to store them. For a large template bank, storing those unimportant bases takes ample memory space even when they are not of any specific use. Only the corresponding singular values are used to evaluate the corresponding $\avgSNRLoss$. For illustrating the wastage of memory space to store unimportant bases, let us consider a template bank with a $10^{5}$ number of template waveforms, such that each waveform has $10^{6}$ time-stamps. Then the required space to store the whole template matrix is $200 \,\GB$. Now for such a template bank, if hypothetically, one considers that the required number of essential basis vectors $\ell = 10^{4}$, then, the memory needed for storing the unimportant ($2N_{T} - \ell = 19\times 10^{4}$) basis vectors is $190 \,\GB$. For this specific example, $95 \%$ of the memory is unnecessarily occupied by those bases which are not used in the $\SVD$ based matched filtering scheme. Also, if only $5\%$ bases are required for the matched filtering calculation, then using $\SVD$ decomposition, one should compute those many bases. In this way, computation costs can be further reduced. However, in this method, the whole set of bases needs to be computed to obtain a $\avgSNRLoss$.  Therefore, although it is not optimal to compute all the bases computationally, it is essential to calculate $\avgSNRLoss$ and obtain a corresponding $\ell$. Thus in the case of the $\SVD$ based matched filtering strategy, one has to compute all the basis vectors knowing the facts that $80-95 \, \%$ of the bases are useless and need to be discarded from the final list. 

It is a real challenging problem constraining the limited computational resources. Ideally, one has to pre-defined a $\avgSNRLoss$ and compute a fixed set of bases that can satisfy it. In this way, the utilization of computational resources will be optimized. However, this kind of reverse calculation is not feasible using $\SVD$ factorization. Addressing this problem using truncated-$\SVD$ set-up is hard. This crucial problem has to be sorted out unless the computational resources will be misapplied for a large template bank. As a result, designing the real-time detection pipeline can not be succeeded. Hence, though the $\SVD$ based matched-filtering approach is the best-known approach for low-latency $\CBC$ searches, it has some specific limitations for a large template bank, as mentioned above. One can summarize the limitations as follows:
\begin{enumerate}
\item Large template matrix can not be decomposed using $\SVD$ due to the high computational cost of decomposition. 
\item For a large template matrix, the reconstruction cost is the dominant cost. Hence, only reducing the filtering cost is not sufficient. 
\item Performing entire $\SVD$ and then choosing $\ell$ bases to attain predefined $\avgSNRLoss$ is inefficient.
\end{enumerate}
To handle the first two problems, the current low-latency $\CBC$ search pipeline ($\GstLAL$) divides the full bank into a group of sub-banks, and for each sub-bank, the $\SVD$ based matched filtering is performing independently. However, this is a sub-optimal solution as an optimal way of dividing the whole bank into sub-banks is an open problem. Secondly, dividing into sub-banks may cause for losing the linear dependency between the nearby waveforms. As a result, it increases the number of essential basis vectors for a fixed $\avgSNRLoss$, which indeed increases the filtering cost.  
In our previous work \cite{kulkarni2019random}, we have computed $\beta$, the ratio of the number of bases summed across all the sub-banks to the number of basis from the $\SVD$ factorization of the full bank. In Figure-$1$ \cite{kulkarni2019random}, we have plotted $\beta$ against $\avgSNRLoss$ for six different template bank. The figure shows that the number of important bases for a whole template bank is less compared to the sum of the bases considering all the sub-banks for a fixed $\avgSNRLoss$. The value of $\beta$ increases with the increasing size of the template bank. This is because the computation of a global set of the basis for a whole template bank is more relevant than computing a local set of bases based on the different sub-banks of the entire template bank as the number of the basis for the previous case is less (see Figure-$1$ of \cite{kulkarni2019random} for more details). Hence, to reduce the filtering cost optimally, one has to compute the global set of bases by directly applying $\SVD$ to the entire template bank. But that is not possible because the computational cost for the factorization for an entire template bank is high. Hence, each sub-bank matrix factorization is the best-known solution for the $\SVD$ based matched-filtering scheme. 
If $\SVD$ factorization for a large template bank is feasible, we may resolve the matched filtering cost optimally compared to the overall filtering cost combining all the sub-banks factorization schemes. However, for a large template bank, matched filtering cost is not only the cost. The reconstruction of the $\SNR$ time-series is also computationally expensive. In fact, for a large template matrix,  reconstruction cost is the dominant cost. Therefore, in the current scheme, to reduce the reconstruction cost, one must divide the full bank into sub-banks. As for each sub-banks, the reconstruction cost is minimal. However, the overall filtering cost is increased due to make a large number of sub-banks. Hence, there is a clear trade-off between the reconstruction cost and filtering cost for a fixed bank. Since the problem of reduction of the filtering cost and the reconstruction cost is complementary. It is not possible to obtain the optimal solution for both issues simultaneously using any matrix factorization (\text{e.g.} $\SVD$) based matched-filtering scheme. The division into sub-banks can reduce the reconstruction cost, whereas it can increase the filtering cost as the number of bases increases by a few factors. The number of basis vectors is directly proportional to the size of the bank. Therefore it can reduce the filtering cost if one computes the global set of bases by factorizing the entire bank instead of obtaining the basis vectors locally from the sub-banks. The optimal time complexity can be related to the division of the bank. Not only that, the whole comparison depends on the number of sub-banks, the size of the template bank, and the number of desired top basis vectors. Therefore, it is crucial to analyze the time complexity calculation in detail for both cases by properly considering all these factors. In section \ref{Subsec:Opt_Temp_div}, we have shown a numerical analysis of the optimized way of division into sub-banks.

In our previous work \cite{kulkarni2019random}, we have shown that the $\RP$ based matrix factorization technique can efficiently compute the set of basis and the coefficient matrix for a large template bank, thus reducing the filtering cost. In this work, we have used a further improved version of the $\RP$ based matrix factorization by which it is possible to identify the value of $\ell$ corresponding to the pre-defined $\avgSNRLoss$. Further, we demonstrate the reduction of the reconstruction cost using the Matching Pursuit ($\MP$) algorithm \cite{pati1993orthogonal}. Finally, we prescribe an algorithm that combines the matrix factorization of a template matrix with a fixed $\avgSNRLoss$ and reduces the reconstruction cost of \SNR \ time-series of each template waveforms. 
\section{Random projection-based template matrix factorization}
\label{Sec:RMF}
In our previous work \cite{kulkarni2019random}, we have demonstrated the utility of the $\RP$-based matrix factorization for the $\GW$ searches from the compact binary coalescence. Consider a template matrix $\mathbf{H}_{2N_{T} \times N_{s}}$, $\exists$ a $\ell$-dimensional dominant subspace $\bar{\mathbf{H}}_{2N_{T} \times \ell} =  \mathbf{H} \, \mathbf{\Omega} \in \mathbb{R}^{\ell}$.
All the row vectors of the template matrix $\mathbf{H}$ have been projected into a $\ell$-dimensional space using an $\RP$ matrix $\mathbf{\Omega}$. Hence the dimension of each of the row vectors reduced from $N_{s} \rightarrow \ell$, where $\ell \ll N_{s}$. Thus, for a large template matrix $\mathbf{H}$, to minimize the factorization cost, it is optimal to use $\bar{\mathbf{H}}$ to obtain the basis of the original matrix as it almost preserves the geometrical structure of the row-space. Hence $\ell$-dimensional representation of the row vectors of $\mathbf{H}$ can be used to compute the basis vectors $\mathbf{Q}$, which are nearly in the same direction as the top-$\ell$ eigenvector directions of the original eigenvector of the row space of $\mathbf{H}$. Concisely, $\RP$ involves taking the projection of a high-dimensional vector to map it into a lower-dimensional space while providing some guarantees on the approximate preservation of pair-wise distance between the vectors. The mere fact that the idea of $\RP$ came directly from $\Johnson$-$\Lindenstrauss$ lemma \cite{johnson1984} guarantees the preservation of the pair-wise distance with a certain accuracy between a set of points which are projected to a lower-dimensional space from a higher dimensional space using $\RP$ operator. The theoretical and practical bound of the distortion factor with an example is shown in the subsection \ref{Subsec:RP_Imp} of section \ref{Sec:appendix}. This $\RP$ based matrix factorization scheme also provides similar factors to obtain rank-$\ell$ approximation ($\mathbf{H}^{(\ell)}$) as using $\SVD$ except for the fact that it factorizes the template matrix into two factors $\mathbf{Q}$ and $\mathbf{B}$ whereas $\SVD$ factorizes it into three different factors. We can use $\mathbf{B}$ as a surrogate template to filter against the data vector $\bm{s}$. Note that the low-rank value ($\ell$) is defined as the dimension of the projected lower-dimensional space; hence, all the template waveforms are projected from $N_{s} \rightarrow \ell$ using $\RP$ operator $\mathbf{\Omega}$ and hence the set of basis $\mathbf{Q}$ has been computed using those template waveforms embedded in a lower-dimensional space. Therefore, the obtained set of basis $\mathbf{Q}$ can be approximated using the top-$\ell$ basis vectors of $\texttt{range}\, (\mathbf{H})$. The coefficient matrix ($\mathbf{B}$) can be obtained by projecting all the row vectors of $\mathbf{H}$ onto the set of basis vectors $\mathbf{Q}$, \text{i.e.} $\mathbf{B} = \mathbf{Q}^{T} \, \mathbf{H}$. Hence, we can approximate $\mathbf{H}_{2N_{T} \times N_{s}} \approx \mathbf{Q}_{2N_{T} \times \ell} \, \mathbf{B}_{\ell \times N_{s}}$. 
Therefore approximated the $\SNR$ time-series ($\rho''_{\alpha}(\Delta t)$) for each of the templates can be computed as follows:
\begin{equation}
\rho''_{\alpha}(\Delta t) = \sqrt{\sum_{\nu = 1}^{\ell}{Q_{(2\alpha-1)\nu} \, \big( \bm{B}_{\nu} \cdot\, \bm{s}^{T} \big) (\Delta t) \, + \, \sum_{\nu =  1}^{\ell} {Q_{(2\alpha)\nu}\, \big( \bm{B}_{\nu}\, \cdot \bm{s}^{T} \big) (\Delta t)}}}
\end{equation}
Based on the projected dimension $\ell$, one can compute the corresponding $\avgSNRLoss$ (See \cite{kulkarni2019random}). Hence, $\mathbf{Q} \, \mathbf{B}$ factorization guarantees  $\|\mathbf{H} - \mathbf{Q \, B}\|_{F} \leq \avgSNRLoss$. It is notable that in this case the obtained $\avgSNRLoss$ can not be predefined before factorization.
However, one can also project the column vectors of $\mathbf{H}$ onto a $\ell$ dimensional space by using a $\RP$ matrix $\mathbf{\Omega}$ of dimension $2N_{T} \times \ell$. In that case, the factorization can be defined as $\mathbf{H}_{2N_{T} \times N_{s}} \approx \mathbf{B}_{2N_{T} \times \ell} \, {\mathbf{Q}^{T}}_{\ell \times N_{s}}$. The details of the algorithm is described in appendix \ref{Subsec:RMF-details}. This kind of factorization is required for the construction of the sparse coefficient matrix using matching pursuit algorithm, described in section \ref{Sec:MP_Algo}. 
Additionally, the $\ell$ dimensional sub-space formation is possible by projecting both the row and column vectors of $\mathbf{H}$ using two $\RP$ matrices. This kind of compression of rows and columns is useful if both the number of rows and columns of the template matrix are large.

\begin{figure}[]
\centering{
\includegraphics[width = .55\textwidth]{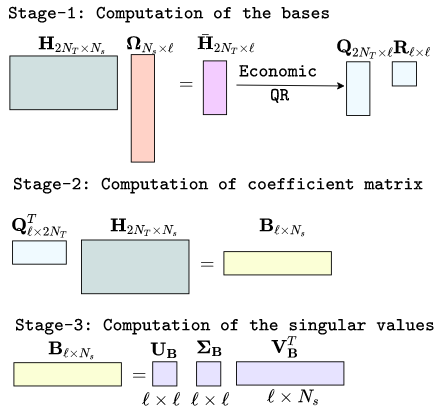}
\caption{The figure shows the pictorial description of the $\RMF$ scheme with a fixed rank $\ell$ for a template matrix $\mathbf{H}_{2N_{T} \times N_{s}}$ (See Alg-\ref{alg:RMF-Basic}). In this scheme, the computation of the orthonormal basis for a template matrix has been done from a lower-dimensional representation of the template matrix, denoted as $\bar{\mathbf{H}}$, which is computed by projecting all the template waveforms to the column vectors of a $\RP$ matrix $\Omega$. The steps are shown in the stage-$1$ of this figure. Since, $\ell \ll N_{s}$, hence the lower-dimensional representational of the template matrix is a much smaller matrix as compare to the original template matrix. Hence, computational expenses can be reduced using this intermediate representation of the template waveforms for a large template matrix. In Stage-$2$, the computation of the coefficient matrix has been shown after obtaining the basis vectors. The Stage-$3$ is an optional step, as it is only useful to obtain the singular value spectra of the coefficient matrix $\mathbf{B}$, which approximates the top-$\ell$ singular values of the template matrix $\mathbf{H}$. The rank-$\ell$ truncation of $\mathbf{H}$ based on $\SVD$ can be equivalently written based on the factors obtained from this $\RMF$ method as $\mathbf{H}_{2N_{T} \times N_{s}} \approx \mathbf{U}_{2N_{T} \times \ell } \, \Sigma_{\ell \times \ell} \, {\mathbf{V}^{T}}_{\ell \times N_{s}}$. Similarly, it can be decomposed as $\mathbf{H}_{2N_{T} \times N_{s}} \approx \mathbf{Q}_{2N_{T} \times \ell} \, \mathbf{B}_{\ell \times N_{s}}$ using $\RMF$. Further, the factors obtained from $\SVD$ and $\RMF$ can be comparable if we decompose the coefficient matrix $\mathbf{B}$ as shown in the Stage-$3$ of this figure. Therefore, one can approximate the left singular basis $\mathbf{U}_{2N_{T} \times \ell} \approx \mathbf{Q}_{2N_{T} \times \ell} \, {\mathbf{U}_{\mathbf{B}}}_{\ell \times \ell}$. Similarly, the top-$\ell$ singular values of $\mathbf{H}$ can be approximated using the singular values of $\mathbf{B}$, \text{i.e.} $\Sigma_{\ell \times \ell} \approx {\Sigma_{\mathbf{B}}}_{\ell \times \ell}$. The right singular basis can also be approximated similar way: $\mathbf{V}_{N_{s}\times \ell} \approx {\mathbf{V}_{\mathbf{B}}}_{N_{s} \times \ell}$. 
}
\label{Fig:RMF-Basic}
}
\end{figure}
The $\RP$ based matrix factorization has the following two crucial benefits:
\begin{itemize}
\item It reduces the time and space complexity of decomposition into factors since the number of dimensions is now quite manageable.
\item Due to the involvement of the simple computational steps, it is easy to make a simple workflow over a high-throughput computing ($\HTC$) environment as well as over distributed-memory architecture, \text{e.g.}, High-performance computing ($\HPC$).
\end{itemize}
Notably, the matrix factorization scheme described in Alg-\ref{alg:RMF-Basic} computes top-$\ell$ basis vectors based on some fixed value of $\ell$. Hence, computation of the corresponding $\avgSNRLoss$ is only possible after obtaining the top-$\ell$ singular values. As $\avgSNRLoss$ is the only tuning parameter for adjusting the approximated \SNR \ time-series obtained by bases-based matched filtering scheme. Therefore it is crucial to design an algorithm that can provide a set of basis vectors based on a pre-defined $\avgSNRLoss$. $\SVD$ or Alg-\ref{alg:RMF-Basic} based matched filtering scheme is inefficient to do that. Therefore in the next section, we describe an algorithm used to obtain the matrix factors corresponding to a pre-defined $\avgSNRLoss$. Mainly the algorithm is an extended version of Alg-\ref{alg:RMF-Basic} in which some optimal rank-$\ell$ approximation of the template matrix has been computed based on fixed $\avgSNRLoss$ iteratively. Identification of optimal $\ell$ based on a pre-defined $\avgSNRLoss$ is also possible iteratively using $\SVD$ factorization by obtaining $\SVD$ factors in each iteration. But for a large template matrix, where applying $\SVD$ for once is computationally expensive, using $\SVD$ over many iterations is not advisable due to the limitation of the computational resources. However, the randomized Alg-\ref{alg:BRMF} is automated and also easily operative on distributed memory architectures.
\subsection{\RP-based template matrix factorization with a fixed average \SNR-loss }
\label{Subsec:RMF with Error}
From the previous section, it is clear that the standard-$\RMF$ scheme ( described in Figure-\ref{Fig:RMF-Basic}, and also in Alg-\ref{alg:RMF-Basic}) only provides a predefined number of basis vectors, and hence there is no such amenability on the approximated error. Therefore, to control the corresponding error due to the factorization, Alg-\ref{alg:RMF-Basic} needs to be modified. Due to this different way of defining the problem statement, $\RMF$ algorithms ~\cite{martinsson2016, halko2011, gu2016} can be classified into the following two classes. 
\begin{enumerate}
\item {\color{blue}{Fixed-rank based factorization}}:
The user provides the predefined rank $\ell$ to obtain the matrix factors. It means that the number of required basis vectors is already fixed before performing the matrix factorization scheme. Hence, the dimension of the projected space is already predefined, and one can do the matrix factorization in that specific dimensional space only. Therefore, the tolerance error due to the factorization will depend on the user-defined rank $\ell$. Alg-\ref{alg:RMF-Basic} employs this kind of factorization, where the user-defined rank of the matrix is $\ell$, and all the higher dimensional vectors are projected onto $\ell$ dimensional space using the $\RP$-operator. 
\item {\color{blue}{Auto-rank based factorization}}:
For this class of algorithms, the numerical rank of a data matrix's approximated factors is not decided beforehand. The algorithm is designed in such a way that the rank corresponding to a specific preset precision error can be revealed automatically. These kinds of algorithms are greedy by nature, wherein the initial guess about the rank has to be made. It is better to start with a small value as a guess of the rank and compute the factorization based on that rank by projecting all the vectors onto that projected space. One can continue the process until the precision error matches the predefined error. This is a kind of block-wise iterative matrix factorization scheme in which, in every iteration, we are obtaining some essential basis and computing the precision error to decide how many basis vectors are sufficient to get the desired precision error. Auto-rank-based factorization can automatically determine the desired rank based on the constraint on the error tolerance. Here, for a template matrix factorization, we have set $\avgSNRLoss$ as the error-tolerance.  
\end{enumerate}

\begin{algorithm}[H]
\DontPrintSemicolon
\SetAlgoLined
\KwIn{Template Matrix \{$\mathbf{H} \in \mathbb{R}^{2N_{T} \times N_{s}}: 2N_{T} \leq N_{s} $\}.}
\KwOut{$\mathbf{Q}_{2N_{T} \times \ell }$, $B_{\ell \times N_{s}}$}
\For{i = 1,2,3, $\cdots$}{
$\mathbf{\Omega}_{i} \in \mathbb{R} ^{N_{s} \times b}  \; : \mathbf{\Omega}_{i} \; \in \mathcal{N}\big(0,1\big) \;$  \;
$\bar{\mathbf{H}} = \mathbf{H} \, \mathbf{\Omega}_{i}$ \atcp{\text{random sampling of column space of} $\mathbf{H}$} \;
$\mathbf{Q}_{i} = \textrm{qr} \, \big (\mathbf{Q}_{i} - \sum_{j = 1}^{i-1}{\mathbf{Q}_{j} \, {\mathbf{Q}_{j}}^{T} \, \mathbf{Q}_{i}}\big )$  \atcp{\text{Re-orthogonalization}} \;
$\mathbf{B}_{i} = {\mathbf{Q}_{i}}^{T} \, \mathbf{H}$ \;
$\mathbf{H} = \mathbf{H} - \mathbf{Q}_{i} \,  \mathbf{B}_{i}$  \atcp{\text{Null space projection}} \;
\If{$\|\mathbf{H}\|_{F} \geq \langle \frac{\delta \rho}{\rho}\rangle$}{$ \mathbf{Q} = \big[\mathbf{Q}_{1}|\mathbf{Q}_{2}|\mathbf{Q}_{3}|\cdots|\mathbf{Q}_{i} \big]$ \atcp{\text{Column-wise stacking}} \;
$\mathbf{B} = \big[{\mathbf{B}_1}^{T}|{\mathbf{B}_{2}}^{T}|\cdots|{\mathbf{B}_{i}}^{T} \big]$ \atcp{Row-wise stacking}\;
}
}
\caption{\RMF \ with a fixed error-$\epsilon$( Block-wise \RMF)}
\label{alg:BRMF}
\end{algorithm}

Alg-\ref{alg:BRMF} \cite{gu2016} employs the $\RP$ of the row vectors iteratively to obtain the factors within a fixed error bound to determine the approximated numerical rank of a data matrix automatically. It is inspired by the column pivoting Gram-Schmidt scheme and combines this scheme along with the random sampling (projection), the blocking to obtain the $\mathbf{Q} \, \mathbf{B}$ factorization. 
Instead of taking the random projection of the row vectors into $\ell$-dimensional subspace directly, a few numbers of blocks $b < \ell$ of $\RP$ matrix have been generated. Firstly, the rows of the template matrix have been projected onto a $b$ dimensional space, and then partial $\QR$ decomposition has been used to obtain the orthogonal basis of that specific ($b$ dimensional) subspace. It is clear that for the first iteration the number of basis vectors will be $b$ and the corresponding $\mathbf{Q}^{(i)} \, \mathbf{B}^{(i)} : i = 1, 2, 3, \cdots,$ represents a $b$ rank approximation of $\mathbf{H}$. After obtaining $\mathbf{Q}^{(i)}$ and $\mathbf{B}^{(i)}$, it is easy to verify the corresponding $\avgSNRLoss$. 
The process continues until it converges to a pre-defined $\avgSNRLoss$. Iterative improvement in the set of orthogonal basis vectors 
($\mathbf{Q} = \big[\mathbf{Q}^{(1)}|\mathbf{Q}^{(2)}|\cdots|\mathbf{Q}^{(i)}\big]$) and dense skeleton matrix ($\mathbf{B} = \big[\mathbf{B}^{(1)}|\mathbf{B}^{(2)}|\cdots|\mathbf{B}^{(i)}$) has been taken into account until it reaches the desired accuracy. By choosing a suitable value of $\avgSNRLoss$, the factorization can be utilized to reveal the value of $\ell$ which provides an optimal rank-$\ell$ approximation of $\mathbf{H}$. 

The $\RP$ matrix $\mathbf{\Omega}_{N_{s} \times \ell} = \Big[\mathbf{\Omega^{(1)}_{N_{s} \times b}}|\mathbf{\Omega^{(2)}_{N_{s} \times b}}|\cdots|\mathbf{\Omega^{(p)}_{N_{s} \times b}}\Big]$ can be thought of as a collection of $p = \lceil \ell/b \rceil$ disjoint sub-random matrices, where the dimension of each sub-random matrices is $N_{s} \times b$. Therefore it turns out to be optimally finding out of the value of $p$ corresponding to a fixed $\avgSNRLoss$. Each of these small blocks can be used to project the template matrix $\mathbf{H}$ into $b$ dimensional space ($\bar{\mathbf{H}}_{2N_{T} \times b} = \mathbf{H}\, \mathbf{\Omega}$). As the dimension of the projected matrix $\bar{\mathbf{H}}$ is small, thus the computation cost for obtaining $b$ number of orthonormal column of $\mathbf{Q}$ using a standard $\QR$ decomposition will be inexpensive. The corresponding $b$ rows of the coefficient matrix $\mathbf{B}$ can also be computed using $\mathbf{Q}$. After getting the factors, the template matrix $\mathbf{H}$ is updated by projecting it out perpendicular to the $b$ basis vectors, described in step-$7$ of Alg-\ref{alg:BRMF}. 

This step is computationally expensive for a large template matrix, as in each iteration, it is required to access the whole template matrix. Thus, we can do this step optimally in a distributed architecture. It is an essential step as it updates the relative error in the matrix approximation in terms of the Frobenius norm. The latter can be easily related to $\avgSNRLoss$. At this point, we need first to check either the target accuracy is reached or not. If not, another pass is made through the updated $\mathbf{H}$ using subsequent blocks of the $\RP$ matrix. It is expected, after a few iterations, one may achieve the optimal rank $\ell$.
While the block-projection scheme described above can lead to computational advantages, it can also lead to the aggregation of round-off errors to compute the basis. Also, a block of basis vectors from a specific iteration is generally not orthogonal to the block of basis vectors obtained from another iteration. Therefore, to ensure the orthogonality of the block of the basis vectors with previously obtained basis vectors, we need to incorporate a Gram-Schmidt like re-orthogonalization procedure to construct the final set of orthonormal basis vectors $\mathbf{Q}$. This operation is shown in step $5$ of Alg-\ref{alg:BRMF}. This re-orthogonalization step is again computationally expensive. Hence for a large template matrix, this step needs to be performed in a distributed way.

\begin{figure}[]
\centering{
\includegraphics[width =.55\textwidth]{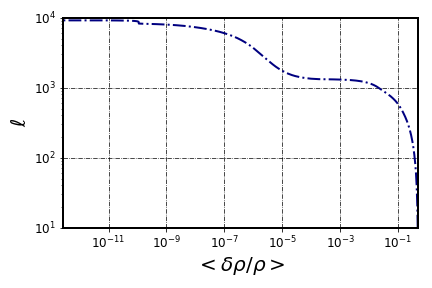}}
\caption{The figure shows the estimated number of basis corresponding to a fixed value of $\avgSNRLoss$. The estimation of basis vectors has been computed using the block-wise $\RMF$ Alg-\ref{alg:BRMF}. The profile shows that number of required basis increases as the value of $\avgSNRLoss$ gets smaller, which is expected. A template matrix $\mathbf{H}$ of size $(2 N_{T} × N_{s}) \equiv 9130  \times 65536$ is used in this analysis. This  template bank has been constructed using non-spinning template waveform for component mass parameters
$\big(m_{1, 2} \big)$ in the range $2.5 M_{\odot} \leq m_{1}, m_{2} \leq 17.5 M_{\odot}$. In each iteration, a block of size $b = 10$ is used to project the template waveform into the lower dimensional space. In every iteration, the average \SNR \ loss is calculated as $\avgSNRLoss \leq \frac{\|\mathbf{H} - \mathbf{H}^{(\ell)} \|_{F}}{\|\mathbf{H}\|_{F}}$. The computation of the Frobenius norm of the original template matrix ($\|\mathbf{H}\|_{F}$) has been pre-computed, and Frobenius norm of $\|\mathbf{H} - \mathbf{H}^{\ell} \|_{F}$ is computed in each iteration. The computation of $\mathbf{H} - \mathbf{H}^{(\ell)}$ is done in step-$8$ of Algorithm-\ref{alg:BRMF}.  }
\label{Fig:RMF-fixed-error}
\end{figure}
\section{Construction of sparse coefficient matrix using Matching pursuit algorithm}
\label{Sec:MP_Algo}
We already have seen that the rows of the template matrix $\mathbf{H}$ can be represented as the linear combination of the set of basis vectors, which can be obtained from $\SVD$ or $\RMF$. If we consider top-$\ell$ basis vectors, then the $\SVD$ decomposition of the template matrix can be shown as follows:
\begin{equation}
\mathbf{H}_{2N_{T} \times N_{s}} \approx \mathbf{U}_{2N_{T} \times \ell} \, \mathbf{\Sigma}_{\ell \times N_{s}} \, {\mathbf{V}^{T}}_{\ell \times N_{s}}
\label{Eq:SVD-Decom}
\end{equation}
It is clear that for the case of $\SVD$ decomposition, $\mathbf{U}$ and $\mathbf{\Sigma}$ together represent a coefficient matrix of dimension $2N_{T} \times \ell$.
Similarly, one can obtain $\RMF$ of $\mathbf{H}$ as follows:
\begin{equation}
\mathbf{H}_{2N_{T} \times N_{s}} \approx \mathbf{B}_{2N_{T} \times \ell} \, {\mathbf{Q}^{T}}_{\ell \times N_{s}}
\label{Eq:RMF-Decom}
\end{equation}
Therefore in generic notation, $H^{\alpha}$ can be written as a combination of the $\alpha^{\th}$ coefficient vector and $\ell$ number of basis vectors. Generally, these coefficient vectors (computed using $\SVD$ or $\RMF$) and indeed the coefficient matrix is a dense matrix, which implies that all the $ \mathcal{O}(N_{T} \, \ell)$ numbers of elements of coefficient matrix are non-zero. This implies that all these weights are important for the construction of the \SNR \ time-series. Therefore, further reduction of the reconstruction cost is not possible. The only possible way to reduce the reconstruction cost is to obtain a sparse coefficient matrix in which every row has a, let say, $k$ number of non-zero weights, where $k < \ell$. For dense coefficient matrix obtained from the matrix factorization scheme, the reconstruction cost is $2N_{T} \, \ell \, N_{s}$ floating-point operations. But if we consider a sparse coefficient matrix, then the reconstruction cost becomes $2N_{T} \, k \, N_{s}$. Hence, the reconstruction cost can be reduced by order of $\ell / k$. Note that we need to choose such a value of $k$ for which the error in the reconstruction $\SNR$ time-series is negligible. One can think of making it sparse by converting some of these non-zero coefficients into zero directly. One possible way is to make the last $\ell-k$ number of weights equal to zero directly. Since all $\ell$ basis vectors are important; hence, all the corresponding weights of the basis have the same importance. Therefore directly converting the last $\ell-k$ number of weights into zero increases the waveform reconstruction error, and indeed it has an immense effect on the reconstruction of the $\SNR$ time-series. Alternatively, one can sort out the coefficients and then convert the last $\ell-k$ coefficients into zero. But this way of transforming from dense to sparse does not work as the effect reflects the reconstruction accuracy of the waveforms. Hence the transformation from the dense to a sparse coefficient matrix by making some non-zero weights into zero-weights is directly is not suggestive. It is essential to assign a new value for the rest of the non-zero weights in such a way that these updated non-zero weights can preserve the length of each template waveforms (\text{i.e.}, the rows of $\mathbf{H}$) with high accuracy. This kind of transformation from dense to sparse coefficients by updating the weights corresponding to each basis vector can be done using the $\MP$ algorithm \cite{pati1993orthogonal}.
This section investigated the possibility of constructing the sparse coefficient matrix using the $\MP$ algorithm. For this purpose, we need a set of basis vectors that can be obtained by factorized template matrix $\mathbf{H}_{2N_{T} \times N_{s}}$ using $\RMF$ or $\SVD$. After factorization, we only choose top-$\ell$ basis vectors for the representation of the waveforms. Hence, the dense coefficient matrix is of dimension $2 N_{T} \times \ell$. if we want to replace the dense coefficient matrix with a sparse one, that implies we want to fix a non-zero number of weights in each row vector as $k$, where $k < \ell$. Then we want to find a new coefficient matrix ($\bar{\mathbf{B}}$, in case of \RMF) which has the same dimension ($2N_{T} \times \ell$) as the previous dense coefficient matrix, but the number of non-zero elements becomes $k$ for each row. If $k < \ell$ and the corresponding reconstructed row vectors of the template matrix can be approximated with high accuracy, we can replace the dense coefficient matrix with the sparse one. The finding of the new coefficients using the $\MP$ algorithm is an optimization problem that can be defined as follows:

\begin{equation}
\begin{aligned}
& \underset{k}{\text{minimize}}
& & \|\bm{x}^{\alpha} \|_{0} , \; \alpha = 1, \ldots, 2N_{T} \\
& \text{subject to}
& & \tilde{\mathbf{H}}^{\alpha} = \bm{x}^{\alpha} \,\mathbf{Q}^{T}.
\end{aligned}
\label{Eq:Optz1}
\end{equation}
The Eq.\ref{Eq:Optz1} represents a sparsity constraint-based optimization problem, where we want to approximate the row vectors of $\mathbf{H}$ with a specific predefined number of sparse coefficients $k$. In this optimization problem, we have assumed that every row vector can be approximated with the same number of non-zero components $k$. One can also choose a different value of $k$ for approximating the different row vectors. Note that, for any condition, the optimal value of $k$ should be computed in a such a way that the approximated row vector of $\mathbf{H}$ can follow $\|\mathbf{H}^{\alpha} -\tilde{\mathbf{H}}^{\alpha}\|_{2} < \delta, \; \alpha = 1, \ldots, 2N_{T}$, where $\delta$ is the $\l$ error for the approximation of the row vectors and it should be small.

The above optimization problem can also be redefined as a $\l$-error ($\epsilon$) constraint-based optimization problem where the number of sparse coefficients is evaluated based on some predefined error bound. Hence, it can be considered as follows:

\begin{equation}
\begin{aligned}
& \epsilon = \underset{\bm{x}^{\alpha}}{\min}
& & \|\mathbf{H}^{\alpha} - \bm{x}^{\alpha} \,\mathbf{Q}^{T} \|_{2}, \\
& \text{subject to}
& & \|\bm{x}^{\alpha}\|_{0} \leq k, \; \alpha = 1, \ldots, 2N_{T}.
\end{aligned}
\label{Eq:Optz2}
\end{equation}
Alternately, we can specify $\epsilon$, the upper bound on the desired target error, and try to minimize $k$, the sparsity, subject to this constraint. 
Both the optimization problems defined in Eq.\ref{Eq:Optz2}, or the alternative, are non-convex optimization problems. In fact, it is a $\NP$-hard problem. It can be sub-optimally solved using the $\MP$ algorithm \cite{pati1993orthogonal}, which is an iterative procedure to obtain the re-weighted non-zero coefficients. It finds each element of a coefficient vector in the step-by-step iterative process. Given a basis $\mathbf{Q}$ and a row vector $\mathbf{H}^{\alpha}$, first fixed initial residual ($\bm{r}$) as $\bm{r} = \mathbf{H}^{\alpha}$ and choose an unselected basis vector $\mathbf{Q}^{(i)}$ from the set of basis $\mathbf{Q}$ and recalculate the residual as $\bm{r} = \mathbf{H}^{\alpha} - \bm{x} \, \mathbf{Q}^{(i)}$. The procedure needs to be repeated until either $\|\bm{x}\|_{0} \leq k$ or $\|\bm{r} \|_{2} \leq \epsilon$ is reached. 

In general, for the $\MP$ algorithm, the Fourier basis, the Haar basis can be used to obtain the sparse coefficient matrix. However, as we already have a set of basis vectors obtained from $\SVD$ or $\RMF$, we can use these set of basis vectors directly for the computation of the sparse coefficient vectors. Hence, this sparse coefficient vectors construction method using $\MP$ is easily fitted with the $\SVD$ or $\RMF$ based match filtering scheme. Here we have used a set of basis vectors obtained from $\RMF$ as an input to the $\MP$ algorithm. For optimal computational cost reduction using the $\RMF$ based matched filtering scheme, we have combined the $\RMF$ method with the $\MP$ algorithm. For a fixed $\ell$, we have demonstrated the proposed $\RMF$ algorithm with sparse coefficients obtained from $\MP$ in Alg-\ref{RMF_MP}.

\begin{algorithm}[H]
\label{RMF_MP}
\DontPrintSemicolon
\SetAlgoLined
\KwIn{Template Matrix \{$\mathbf{H} \in \mathbb{R}^{2N_{T} \times N_{s}}: 2N_{T} \leq N_{s} $\}, $\ell$, $k$}
\KwOut{$\mathbf{Q}_{2N_{T} \times k }$, $\mathbf{\tilde{B}}_{k \times N_{s}}$}
$\mathbf{\Omega} \in \mathbb{R} ^{\ell \times 2N_{T}}  \; : \mathbf{\Omega}_{ij} \; \in \mathcal{N}\big(0,1\big) \;$  \;
${\bar{\mathbf{H}}}_{\ell \times N_{s}} = \mathbf{\Omega}_{\ell \times 2N_{T}} \, \mathbf{H}_{2N_{T} \times N_{s}}$ \;
$\mathbf{Q}_{\ell \times N_{s}} = \textrm{qr}(\bar{\mathbf {H}})$ \;
\For{$\alpha$ = 1,2,3, $\cdots$, $2N_{T}$}{
$\bm{\tilde{b}}^{\alpha} = \MP \big( H^{\alpha}, \mathbf{Q}^{T} \big)$ \atcp{Using Matching Pursuit algorithm}
${\mathbf{\tilde{B}}}_{2N_{T} \times \ell} = [\bm{\tilde{b}}^{1}|\bm{\tilde{b}}^{2}|\cdots|\bm{\tilde{b}}^{\alpha}]$ \atcp{Row-wise stacking}\;
}
\caption{$\RMF$ with sparse coefficient}
\label{alg:RMF-MP}
\end{algorithm}

\begin{figure}[htbp]
\centering{
\includegraphics[width = .9\textwidth]{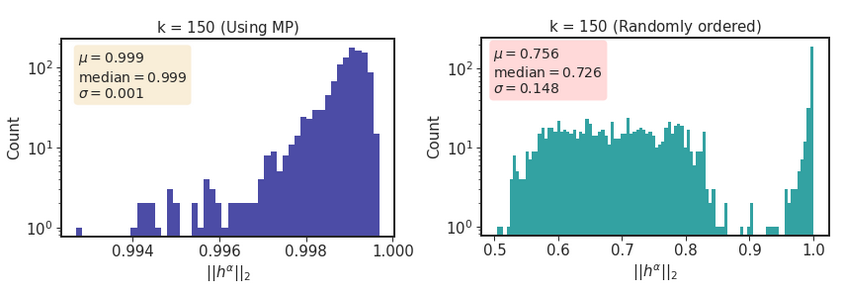}
\caption{The left panel of the figure shows the distribution of the norm ($\l$) of the reconstructed waveform using a sparse coefficient vector obtained from the $\MP$ algorithm, and the bases are computed using $\RMF$. The size of the template matrix is $1162 \times 16384$. $\ell = 200$ number of basis chosen based on the $\avgSNRLoss = 2 \times 10^{-4}$ and out of $200$ non-zero coefficients, $k = 150$ coefficients are consider as non-zero for the formation of the sparse coefficient vectors. The figure clearly shows that using sparse coefficient vectors; we can also preserve the norm of the template waveform within an error of $10^{-2}$. }
\label{Fig:Hist_Coeff_MP}
}
\end{figure}
We now present the efficiency of the $\MP$ algorithm to obtain the sparse coefficient vectors for each template waveform (\text{i.e.}, the row vectors of $\mathbf{H}$). As we used normalized template waveforms (\text{i.e.}, the $\l$-norm is unit), it is expected, after reconstruction of the template waveforms using sparse coefficients obtained from the $\MP$ algorithm and a set of top-$\ell$ basis vectors obtained from $\RMF$ can also approximate the norm of each template waveform with high accuracy nearly to the unit. Therefore, we have performed a simulation and computed the $\l$-norm of reconstructed rows of $\mathbf{H}$ using $k$-sparse coefficient vector obtained using $\MP$ algorithm and a set of top-$\ell$ basis vectors obtained from $\RMF$.
%
Figure-\ref{Fig:Hist_Coeff_MP} shows the histogram of the $\l$-norm of each row vector after reconstruction using Alg-\ref{alg:RMF-MP}. For this example, we consider a template bank $\mathbf{H}$ containing $N_{T} = 581$ templates covering the component mass space:  $5 \leq m_{1,2}/M_{\odot} \leq 15$ using non-spinning $\texttt{TaylorT4}$ waveforms. Each waveform was taken to be $8$ seconds long, sampled at $2048$ $\Hz$, thereby setting $N_{s} = 16384$. The required number of basis $\ell = 200$ is fixed based on the $\avgSNRLoss = 2 \times 10^{-4}$. The sparsity for each of the coefficient vectors is considered as $k = 150$. It is expected that due to the conversion from the dense coefficient vector to the sparse coefficient vector, the $\l$-norm of the reconstructed rows can not be of exact unit magnitude. But from Figure-\ref{Fig:Hist_Coeff_MP}, it is clear that the maximum error in the norm is $10^{-2}$. Hence the sparsification of the coefficient vector can preserve the norm of each of the row vectors with high accuracy. Therefore, we can use the sparse coefficient matrix for the computation of the $\SNR$ time-series. Hence, for this specific example, using the sparse coefficient matrix, we reduced the reconstruction cost by $25\%$ of the previous reconstruction cost considering the dense coefficient matrix. This is because we choose a value of $k = \frac{3}{4} \, \ell$. 
\begin{figure}[htbp]
\centering{
\includegraphics[width = .75\textwidth]{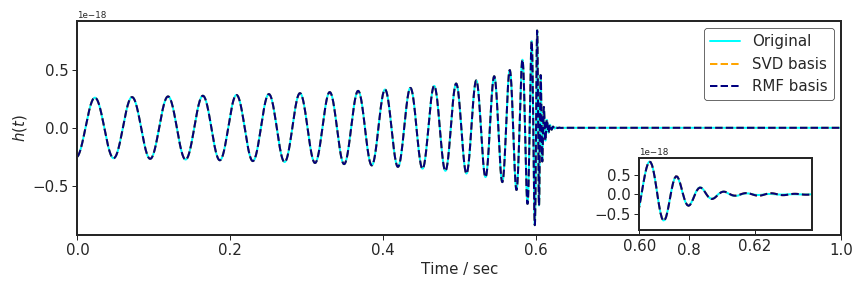}
\caption{The figure shows the reconstructed waveform corresponding to a mass component $m_{1}, m_{2} = 36 \,, 40 \, M_{\odot}$. We have chosen a set of $2000$ template waveforms from the range of mass components $30$-$50\, M_{\odot}$. Out of $2000$ basis vectors, top-$100$ ($\ell = 100$) basis vectors have been used to reconstruct the waveform. However, for the reconstruction of the waveform, we have not used the coefficient vector obtained from the $\SVD$. We have used the $\MP$ scheme to make the more sparse coefficients. We have used half of the coefficient vector's length (\text{i.e.} out of $100$, \text{i.e.}, $k = 50$) to construct sparse coefficients. The reconstruction is done using top-$100$ $\SVD$-basis and top-$100$ $\RMF$ basis (shown in orange and blue colours). The reconstruction accuracy for both cases is almost the same.
}
\label{fig:Example-GW150519}
}
\end{figure}


\section{Implementation of \RMF \ for \LLOID-type framework}
\label{Sec:LLOID-RMF}
\begin{figure}[htbp]
\centering{
\includegraphics[width =.65\textwidth]{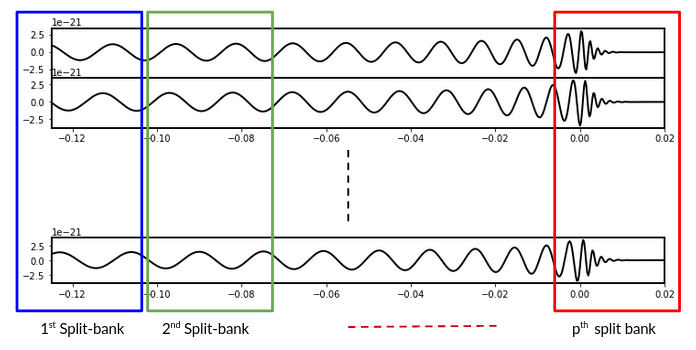}}
\caption{The figure demonstrates the construction of the split banks. For a specific sub-bank, if the waveforms are long (\text{i.e.} the case for $\BNS$, $\NSBH$, and low-mass $\BBH$ systems, we can split every waveform into several parts and stacked them together to construct a split bank. The blue, green, and red boxes describe such a split bank. In this figure, the red box represents a split template matrix containing the merger and ringdown part, whereas the blue and green boxes contain the early inspiral part. Alg-\ref{alg:DrMF_c}, and Figure-\ref{Fig:RMF-columnwise} describe the strategy to combine these split banks to obtain the set of basis vectors for the full one. Alg-\ref{alg:BDrMF_c2} shows the same scheme but with a fixed error instead of a predefined rank $\ell$.  }
\label{Fig:Split-Bank-Demo}
\end{figure}

In the $\GstLAL$ pipeline, a full template bank is divided into several sub-banks. The sub-banks division is done based on the threshold on the chirp-mass and the duration of the template waveforms. The template waveforms from each sub-bank can further be split into several time slices based on different sampling frequency rates \cite{cannon2012toward}. The early inspiral signal can be sampled using a low sampling rate as it enters the detection bandwidth with low frequency. At the same time, merger and ringdown (up-to-last stable orbit) signals contain high-frequency components. Thus required sampling rate for the merger and ringdown will be high as compared to the inspiral part. Also, the time duration for inspiral, merger, and ringdown is different. Due to the low frequency, the inspiral waveform duration is longer compared to the merger, and similarly, the merger waveform has a long duration compared to the ringdown segment. Adapting this idea, any waveforms from a sub-bank can be sliced into early inspiral, late inspiral, ringdown, merger, and sampled with different sampling rates using the down-sample method. For example, The early and late inspiral parts of waveforms can be sampled with a low sampling rate \text{e.g.} $32$-$256$ $\Hz$. The merger and ringdown parts can be sampled with $512$-$2048$ $\Hz$ depending on the duration of the template waveform. Figure-\ref{Fig:Time-Vs-Freq-Evol-NSBH} demonstrates an example of the time-frequency evolution of a waveform from a $\NSBH$ system and how several sampling frequencies are used to represent the early inspiral, late inspiral, merger, and ringdown parts of the waveform. A split bank is defined by combining a specific time slice of all template waveforms in a given sub-bank. Each sub-bank has several split banks. Therefore, instead of computing the basis vectors for a sub-bank, it computes several independent $\SVD$ decompositions based on split-banks. Each split bank decomposition only contributes to calculating a specific fraction of the \SNR \ time-series corresponding to a particular sampling rate.
Further, to obtain the complete $\SNR$ time-series based on a fixed (uniform) sampling rate, one needs to use the $\sinc$ interpolation scheme to upsampled the matched filter output for those time-slices for which the sampling rate is low.  Combining them with the \SNR \ time-series of the other slices of the high sampling rate will provide the full approximated \SNR \ time-series based on the top few basis vectors. The mathematical construction of the $\LLOID$ scheme is described in the Ref. \cite{messick2017analysis}. The approximation of $\SNR$ time-series using $\LLOID$ strategy plays a vital role on a region of $\CBC$ parameter space centered on $\BNS$ masses, for $\BNS$ system, the waveforms are very long, hence required to apply $\LLOID$ scheme for the reduction of the filtering cost. However, the total number of individual $\SVD$ decomposition operations has been increased in that process. Also, an extra computation cost is incurred due to the use of the $\sinc$ interpolation scheme for the upsampling of the partial $\SNR$ time-series.  
This section presents $\RMF$ schemes that can be useful to obtain basis vectors for a split bank. Our definition of a 'split bank' is similar to the description used in a $\LLOID$ framework. However, the main difference in the definition is that we used the fixed (uniform) sampling rate for all the phases (\text{i.e.}, early and late inspiral, merger, and ringdown) of the template waveform. That implies the sampling rate to generate the merger and ringdown phase is the same as the inspiral one. Since our main objective is to obtain common basis vectors combining all the split banks, we need to use the uniform sampling rate for the generation of each time-slices. We have used the finally obtained common set of basis vectors for the computation of the $\SNR$ time-series. As we used a fixed sampling rate, downsampling for each phase and the upsampling step via $\sinc$ interpolation in the $\LLOID$ scheme are easily excluded in our approach to obtain the \SNR \ time-series. Thus, our proposed process is simple and computationally less expensive. For each split bank, the computation of independent basis vectors via $\SVD$ is not required. Hence, we not only reduce the overall cost of performing several $\SVD$ considering all split banks but, further, the cost of downsampling and upsampling have vanished. Figure-\ref{Fig:Split-Bank-Demo} demonstrates a split bank for our purpose. The blue box defines the first split bank, which contains the time-slices of the template waveforms for the early inspiral part. Similarly, the green and the red boxes represent the second and $p^{\th}$ split bank. It is clear from the figure that the red box contains the merger time-slice part only. The proposed algorithm is mainly designed for the handling of a set of long waveforms (\text{i.e.}, for $\BNS$ system). However, it can also be successfully implemented for a group of short (\text{i.e.}, for high mass $\BBH$ system) or medium-duration (\text{i.e.}, for low mass $\BBH$ or $\NSBH$ system) waveforms. In Figure-\ref{Fig:Parameter-space}, we have pictorially shown the applicability of our various $\RMF$ algorithms to the different regimes of the parameter space for a $\CBC$ sources. 

\subsection{Proposed method}
\begin{figure}[htbp]
\centering{
\includegraphics[width =.95\textwidth]{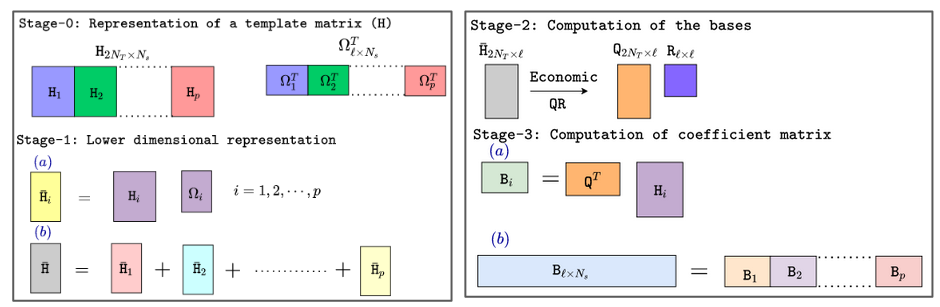}}
\caption{The figure demonstrates the steps involved in the $\RMF$ scheme (See Alg-\ref{alg:DrMF_c} of the appendix) similar to the $\LLOID$-type framework. }
\label{Fig:RMF-columnwise}
\end{figure}

\begin{figure}[htbp]
\centering{
\includegraphics[width =.6\textwidth]{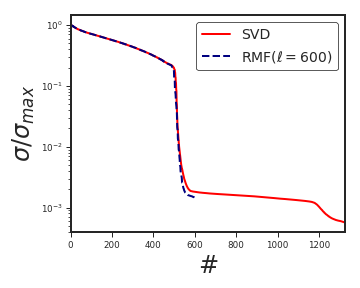}
\caption{The figure shows the comparison of singular values ($\sigma$) for a template matrix $\mathbf{H}$ of size $(2N_{T} \times N_{s}) \equiv 1324 \times 225280$, normalized by the maximum singular value ($\sigma_{max}$) as obtained from $\SVD$ and $\RMF$ algorithm shown in Alg-\ref{alg:DrMF_c}. The $\LLOID$-type framework has been constructed and prescribed $\RMF$ (Alg-\ref{alg:DrMF_c}) is performed in the target dimensions $\mathbb{R}^{\ell}$, where $\ell = 600$. In this example, template bank was constructed using a non-spinning waveform model for component mass parameters ($m_{1,2}$) in the range $1.0 M_{\odot} \leq m_{1}, m_{2} \leq 1.1 M_{\odot}$. The waveforms are generated using the IMRPhenomD waveform model. The lower cut-off frequency for the waveform generation is chosen as $10$ $\Hz$. Therefore the duration of the longest waveform of this bank becomes $110$ sec. Further, the $2048$ $\Hz$ sampling rate is used for the generation of the waveforms. The waveforms are split into $11$ split-banks, and hence the duration of each time-slices becomes $10$ sec. Therefore, each template matrix's size corresponding to the split bank becomes $1324 \times 22528$. As seen here, the top-$\ell$ singular-values obtained by $\RMF$ agree very well with the spectrum obtained by traditional $\SVD$ factorization.}
\label{Fig:RMF-Col-Example-Sig}
}
\end{figure}

\begin{figure}[htbp]
\centering{
\includegraphics[width =.6\textwidth]{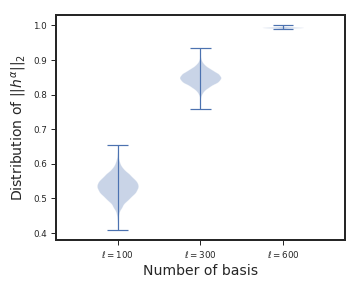}
\caption{The figure shows the distribution of $\l$-norm of the reconstructed template waveform using $\ell = 100, 300, 600$ for the template matrix $\mathbf{H}$ used in Figure-\ref{Fig:RMF-Col-Example-Sig}. It is clear from the figure that the mean of the reconstructed waveforms' norm improves with the increasing value of the rank $\ell$. Similarly, the variance of the norm decreases with a rising value of $\ell$. This plot shows the efficacy of the proposed $\RMF$ scheme shown in Alg-\ref{alg:DrMF_c} as the obtained basis vectors along with the coefficient matrix can reconstruct the waveform with very high accuracy. }
\label{Fig:RMF-Col-Example-dist}
}
\end{figure}

For the description of the scheme shown in Alg-\ref{Fig:RMF-columnwise}, we assumed that Figure-\ref{Fig:Split-Bank-Demo} depicts a bank with longer waveforms. That implies that the column dimension of the constructed template matrix is greater than the row dimension \text{i.e.} $N_{s} \gg 2N_{T}$. 
Let us consider $\mathbf{H} = \{\mathbf{H}_{1} | \mathbf{H}_{2}|\cdots|\mathbf{H}_{i}\} : i = 1, 2, \cdots p$ is a template matrix and $\mathbf{H}_{i}$ represents a split-bank corresponding to the $i^{\th}$ time-slices of the template waveform. The dimension of each $\mathbf{H}_{i}: i = 1, 2, \cdots, p$ becomes $2N_{T} \times \frac{N_{s}}{p}$. 
For simplicity, we have assumed that each split-bank has the same dimension. But, one can also choose a different size for different time-slices. In that case, the column-dimension of the template matrix corresponding to each individual split-bank will be different. However, our approach is independent of the split-banks' column dimension as our final objective is to obtain the rank-$\ell$ factorization of a template matrix by accessing the individual split-bank only. 
Suppose we have only access to each of these $\mathbf{H}_{i}$ independently that implies these template matrices are stored in the independent machines. Thus we want to propose a $\RMF$ scheme which can be useful to obtain a $\ell$-rank factorization of the whole template matrix $\mathbf{H}$ as $\mathbf{Q}_{2N_{T} \times \ell}\,\mathbf{B}_{\ell \times N_{s}}$ using these individual template matrices $\mathbf{H}_{i}$. The designed scheme is shown in Alg-\ref{alg:DrMF_c}. Similar scheme using block-wise $\RMF$ is also shown in Alg-\ref{alg:BDrMF_c2}. The steps involved in this algorithm are also discussed in the Figure-\ref{Fig:RMF-columnwise}.  
In this figure, the stage-$0$ represents a distributed architecture, in which independent machine has individual fraction of the template waveform \text{i.e.} $\mathbf{H}_{2N_{T} \times \frac{N_{s}}{p}}$. Further each machine can generate a specific fraction of the $\RP$ matrix \text{i.e.} ${\Omega_{i}}_{\frac{N_{s}}{p} \times \ell}$ to project a fraction of template waveforms in a lower (say $b$) dimensional space. Hence, the dimension of each row vectors of $\bar{\mathbf{H}}_{i}$ becomes $2N_{T} \times \ell$. This step is shown in the Figure-\ref{Fig:RMF-columnwise}, $\Stage$-$1(a)$. The benefit of this approach is that no need to generate a large $\RP$ matrix $\mathbf{\Omega}_{N_{s} \times \ell}$ to obtain a lower dimensional represent of long template waveforms. Further, ${\mathbf{\Omega}_{i}}_{\frac{N_{s}}{p} \times \ell}$ can be generated independently in each of machines, hence there are zero communication cost is involved. In this way, reduction of the space as well as computational cost for lower-dimensional representation of the waveform can be possible. However, individual lower-dimensional representation of the $\Split$-$\bank$ can not be suitable to obtain the top-$\ell$ basis vectors of the compressed row vector of the full template matrix. Thus to compute the top-$\ell$ basis vectors from the lower dimension representation of the whole template matrix, we need to add all these individual projected components $\bar{\mathbf{H}}_{i} : i = 1, 2, \cdots, p$ together. The addition of these projected components, mathematically shown as $\bar{\mathbf{H}} = \sum_{i = 1}^{p}{\bar{\mathbf{H}}_{i}}$ 
(Shown in $\texttt{Stage}$-$1(b)$. In the next stage ($\stage$-$2(a)$), the top-$\ell$ basis vectors can be computed using $\QR$ decomposition. The next stage is optional and only required if we want to compare the singular values of the original data matrix along with the compressed matrix $\mathbf{B}$. To compute the individual coefficient matrix $\mathbf{B}_{i}$, we need to re-distribute the corresponding fraction of $\mathbf{Q}_{i}$ to each of the machines which already has stored the fraction of the original template matrix. Finally, in $\stage$-$3(b)$, we need to stack them together to get the full coefficient matrix corresponding to the original template matrix. Alternatively, computation of the coefficient matrix can be done using the $\MP$ algorithm; in that case, no need to generate the coefficient matrix $\mathbf{B}$ using the projection operation shown by $\texttt{stage}$-$3$. Figure-\ref{Fig:RMF-Col-Example-Sig} and Figure-\ref{Fig:RMF-Col-Example-dist} described the efficiency of the proposed scheme shown in Alg-\ref{alg:DrMF_c}.  

\subsubsection{\RMF: combining all sub-banks}
\begin{figure}[htbp]
\centering{
\includegraphics[width=.50\textwidth]{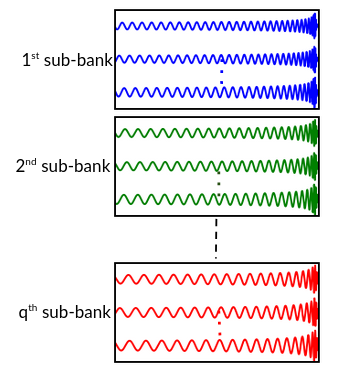}}
\caption{The figure demonstrates the division of a template bank into several sub-banks. The sub-bank division for non-precession waveforms is generally done based on the chirp-mass threshold ($\mathcal{M}$) of the template waveforms. This figure assumes that the waveforms in a template bank are sorted based on the chirp-mass value, and the different color represents different sub-banks. For each sub-bank, we can generate a template matrix. The separate colored box of waveforms represents the sub-template matrix corresponding to each sub-bank. Using $\SVD$ or $\RMF$, one can compute a set of the basis for each sub-banks. However, it is hard to combine all the sub-banks and obtain a global collection of the basis for the whole bank if the bank size is large. However, using our proposed $\RMF$ algorithm, we can combine them and obtain a global set of bases, which can be useful for representing the template waveforms for the whole template bank. The proposed algorithm is shown in Alg-\ref{alg:DrMF_r} and also described using simple steps in Figure-\ref{Fig:RMF-rowwise}. }
\label{fig:Sub-Bank-Demo}
\end{figure}

It is difficult for a huge bank to factorize the full template matrix together in $\SVD$ set-up even in a distributed architecture. Hence, we aim to investigate the whole template matrix's factorization by exploring the $\RMF$ set-up in a distributed manner. Firstly consider that the template matrix $\mathbf{H}$ is too large to store in a single machine, and each machine can store a fixed number of templates. That implies that each machine can contain a specific sub-bank. In Figure-\ref{fig:Sub-Bank-Demo}, we assumed that a template matrix corresponding to a whole template bank is divided into $q$ number of sub-matrices that are equivalent to the fact that the entire bank is divided into $q$ number of sub-banks. We can apply $\RMF$ or $\SVD$ independently to obtain the corresponding basis vectors for each sub-bank. However, we aim to design a scheme using $\RMF$ for getting a standard set of bases by combining all the sub-banks, which is beyond the scope of the current framework of the $\GstLAL$ pipeline. 
\begin{figure}[htbp]
\centering{
\includegraphics[width =.65\textwidth]{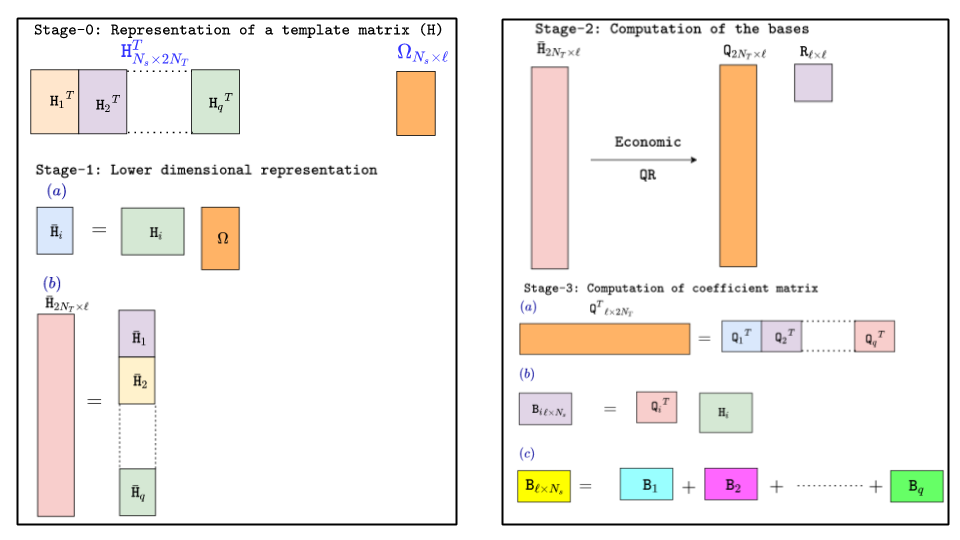}
\caption{The figure demonstrates the steps involved in the $\RMF$ algorithm (See Alg-\ref{alg:DrMF_r}) designed to compute a global collection of bases, combining all the sub-banks.}
\label{Fig:RMF-rowwise}
}
\end{figure}
The problem is similar to the problem defined in Alg-\ref{alg:DrMF_c}. The only difference is that in this set-up, we have a fixed number of time-samples $N_{s}$ for each template waveforms and stored a fraction of the total number of waveforms in a fixed machine, \text{i.e.}, the template matrix corresponding to a sub-bank. The whole scheme can be designed in a distributed set-up where each of these split-banks $\mathbf{H}_{i}$ can be stored in an independent machine. As the entire template matrix is distributed over different devices, it is impractical to perform the $\SVD$ or fixed rank $\RMF$ scheme onto the whole template matrix $\mathbf{H}$ directly. Thus, we have prescribed a scheme using block-wise $\RMF$ by which it is possible to get a rank-$\ell$ approximation of the full template matrix. The details of the algorithm defined in Alg-\ref{alg:DrMF_r}, Alg-\ref{alg:BDrMF_r2}.   

The steps involved in this scheme are shown in Alg-{\ref{alg:DrMF_r}}. This scheme is a distributed version of Alg-{\ref{alg:RMF-Basic}}, where the steps are designed as the same spirit of Alg-{\ref{alg:RMF-Basic}}, and also in each step, simple matrix algebraic operations are adapted. Suppose we have $p$ number of machines and one central node that can communicate between other machines. In Alg- {\ref{alg:DrMF_r}}, the step-$1$ shows the generation of $\RP$ matrix $\mathbf{\Omega}$ of size $N_{s} \times \ell$ in each of the machine. Each machine contains $2{N_{T}}_{b}$ number of rows of the template matrix $\mathbf{H}$. Therefore, the lower-dimensional representation using $\RP$ of these $2{N_{T}}_{b}$ number of rows are shown in step-$2$ by constructing $\bar{\mathbf{H}}_{i}$. If $b$ is small, then the column dimension of these $\bar{\mathbf{H}}_{i}$ matrices are small enough to fit them into local memory ($\RAM$) each of the machines easily. Next, we should transfer all this lower-dimensional representation (\text{i.e.}, in a $\ell$ dimensional space) of the row vectors to the central machine for obtaining the $\ell$ number of basis vectors. Hence, step-$3$ shows the communication between all the individual machines with the central machine to collect all $\ell$ dimensional representation of the rows of $\mathbf{H}$ defining by $\bar{\mathbf{H}}$. After collecting all lower-dimensional representations of the row vectors in the central node, one can perform a $\QR$ decomposition on $\bar{\mathbf{H}}$ to obtain $\ell$ number of basis vectors $\mathbf{Q}$. The step-$4$ illustrates this operation. To compute the coefficient matrix $\mathbf{B}$ in a distributed set-up, again, we need to pass $2{N_{T}}_{b}$ numbers of rows of $\mathbf{Q}$ to each of the machines, which is shown in step-$5$. The step-$6$ showed the computation of $\mathbf{B}_{i}$ of dimension $\ell \times N_{s}$ by projecting the $2{N_{T}}_{b}$ basis vectors onto $2{N_{T}}_{b}$ rows of $\mathbf{H}$ independently in each machine. But the final $\mathbf{B}$ should be a projection of $\ell$ basis vectors onto all the rows of $\mathbf{H}$. Hence after getting $\mathbf{B}_{i}$, one has to send all this $\mathbf{B}_{i}$ to the central node to compute $\mathbf{B}$, which is shown in step-$7$. Note that a total of three communications between the central node and the local machines are required. These are reflected by the steps $3$, $5$, and $7$, respectively. The communication cost for the steps-$3$ and $5$ are $2{N_{T}}_{b} \times \ell$ floating-point operations for each of the local machines. Similarly, for step-$7$, the communication cost is $\ell \times N_{s}$.

\begin{figure}[htbp]
\centering{
\includegraphics[width =.6\textwidth]{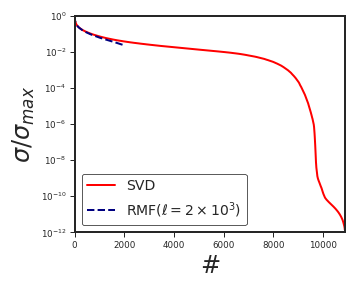}
\caption{The figure shows the comparison of singular values ($\sigma$) for a template matrix $\mathbf{H}$ of size $(2N_{T} \times N_{s}) \equiv 10888 \times 18432$, normalized by the maximum singular value ($\sigma_{max}$) as obtained from $\SVD$ and $\RMF$ algorithm shown in Alg-\ref{alg:DrMF_r}. The prescribed $\RMF$ (Alg-\ref{alg:DrMF_r}) scheme is performed in the target dimensions $\mathbb{R}^{\ell}$, where $\ell = 2000$. In this example, the template bank was constructed using aligned-spin signal model for component mass parameters ($m_{1,2}$) in the range $5.0 M_{\odot} \leq m_{1}, m_{2} \leq 35.0 M_{\odot}$. The range of dimensionless spin magnitude was taken to upto $0.1$. Stochastic template bank placement algorithm is used to generate the template bank. Further, IMRPhenomPv2 signal model is used for the generation of the waveforms. The lower cut-off frequency for the waveform generation is chosen as $30$ $\Hz$. Further, the $2048$ $\Hz$ sampling rate is used for the generation of the waveforms. The waveforms are split into $11$ sub-banks \text{i.e.} $\mathbf{H} = |{\mathbf{H}_{1}}_{1000 \times 18432}|{\mathbf{H}_{2}}_{1000 \times 18432}|\cdots|{\mathbf{H}_{11}}_{888 \times 18432}|$. As seen here, the top-$\ell$ singular-values obtained by $\RMF$ agree very well with the spectrum obtained by traditional $\SVD$ factorization.}
\label{Fig:RMF-Row-Example-Sig}
}
\end{figure}
\begin{figure}[htbp]
\centering{
\includegraphics[width =.6\textwidth]{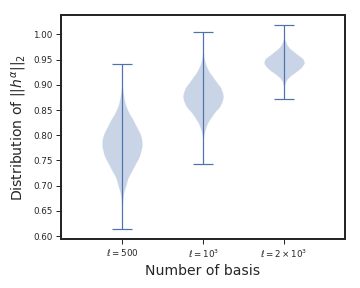}
\caption{The figure shows the distribution of $\l$-norm of the reconstructed template waveform using $\ell = 500, 10^{3}, 2 \times 10^{3}$ for the template matrix $\mathbf{H}$ used in Figure-\ref{Fig:RMF-Row-Example-Sig}. The mean of the reconstructed waveforms' norm improves with the increasing value of the rank $\ell$. Similarly, the variance of the norm decreases with a rising value of $\ell$. This plot shows the efficacy of the proposed $\RMF$ scheme shown in Alg-\ref{alg:DrMF_r} as the obtained basis vectors along with the coefficient matrix can reconstruct the waveform with high accuracy.}
\label{Fig:RMF-Row-Example-dist}
}
\end{figure}
\subsection{Numerical Simulation}
\label{SubSec:Num-sim}
\begin{figure}[htbp]
\centering{
\includegraphics[width = .55\textwidth]{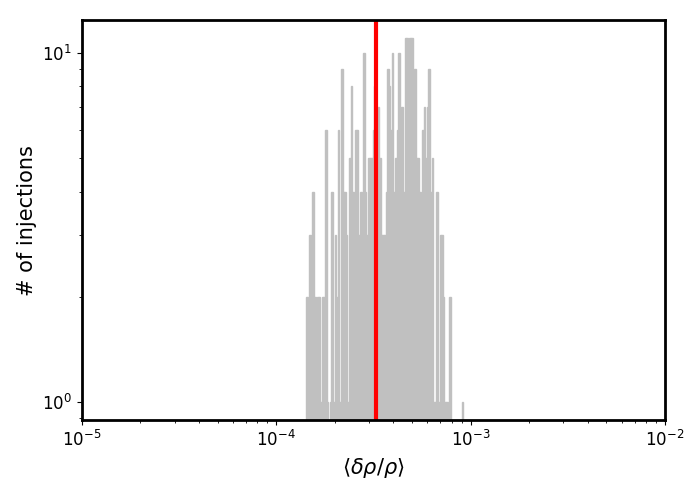}
\caption{It shows the distribution of $\avgSNRLoss$ obtained from a Monte-Carlo injection study. This experiment is similar to the numerical study carried out in \cite{cannon2010}. Here, we have applied $\RMF$-based factorization of the template banks. Further, the $\SNR$ time-series reconstruction is done based on the sparse-coefficient obtained using the $\MP$ algorithm. This study is reflecting the performance of Alg-\ref{alg:RMF-MP}, in which $\ell = 200$ and $k = 150$ have been chosen.  
}
\label{Fig:MC_RMF_MP}
}
\end{figure}

Here, we have done a Monte-Carlo simulation to reconstruct the \SNR \ time-series for a $500$ randomly sampled injection from a specific parameter space. We consider a template bank $\mathbf{H}$ which contains $N_{T} = 581$ templates covering the component mass space: $5 \leq m_{1,2}/M_{\odot} \leq 15$ using non-spinning $\TaylorT4$ waveforms. Each waveform was taken to be $8$ seconds long, sampled at $2048$ $\Hz$, thereby setting $N_{s} = 16384$. Hence the size of the template matrix is $1162 \times 16384$. The required number of basis $\ell = 200$ is fixed based on the $\avgSNRLoss = 2 \times 10^{-4}$. First, we have computed the top-$200$ basis vectors, and we have used these bases for obtaining a sparse coefficient for each of the template waveforms using the $\MP$ algorithm. For sparse representation, we have chosen $k = 150$ non-zero elements out of $200$, for each coefficient vector. The simulation result is shown in Figure-\ref{Fig:MC_RMF_MP}.
\section{Discussion and Conclusion}
This work demonstrated the practical implementation strategy for computation of matched filter output between data and template waveforms combining block-wise $\RMF$ and $\MP$ algorithm. 
Block-wise $\RMF$ can be used to compute an essential set of basis vectors, whereas $\MP$ can be used for obtaining sparse coefficients of the template waveforms. Further, this work prescribed some advanced $\RMF$ algorithms (block-wise) for computing a basis for a large bank by combining all the sub-banks and similarly for long waveforms incorporating all split banks. In addition, we have designed an alternative and computationally efficient framework similar to the $\LLOID$ based on prescribed algorithms and explored several regions ($\BNS$, $\BBH$, $\NSBH$) of the parameter space to demonstrate the efficiency of our framework.
The work presented here has mainly three aspects. 
\begin{enumerate}
\item The designed framework is efficient to obtain the required number of basis vectors at a fixed average \SNR \, loss.
\item Secondly, block-wise $\RMF$ algorithms with a combination of $\MP$-based sparse coefficient construction can also address the issue of reducing the reconstruction cost of computing matched-filter output. It is notable that the $\MP$-based sparse coefficient construction scheme is entirely independent of the $\RMF$ algorithm and can be easily adaptable along with other basis finding methods, \text{e.g.}, $\SVD$. Hence, the sparse coefficient computation using $\MP$ can be directly applicable to the $\GstLAL$ pipeline in the current framework if someone wants to reduce the cost of reconstruction. The highlights of the work are as follows:  
\item  Here, we have demonstrated the advanced block-wise $\RMF$ Algorithms as an alternative to the $\LLOID$ framework. This scheme can split the waveform into several parts, similar to the waveform splitting in the $\LLOID$ method. However, the fundamental difference between the two approaches is that in the $\LLOID$ scheme, downsampled is used for the different parts ( \text{i.e.}, early inspiral, late inspiral, merger, and ring-down) of the waveform to reduce the size of the split-bank. In our approach, we have used a uniform sampling frequency throughout every part of the waveform. Due to the use of downsampling, the process of obtaining the whole $\SNR$ time series in the $\LLOID$ framework is complicated, computationally expensive, as after receiving the $\SNR$ from each split part, upsampling is also required before combining them to get the full-one. Our scheme is simple, as we use the same sampling frequency, so no need to upsample the obtained $\SNR$ from each split part. The use of downsampling-based representation can be possible for our proposed setup. In that case, we need to compute the basis vectors of this split-part separately, similar to the current framework. Since our objective is to obtain one set of bases, combining all the split parts. Hence, our designed algorithms are solely made to get a standard set basis using all the divided parts. Incorporating $\RMF$ in a $\LLOID$ framework is a straightforward task in which we need to replace the $\SVD$ method with $\RMF$.  We will explore the current $\LLOID$ framework of $\GstLAL$ using the $\RMF$ setup in the upcoming work.    
\end{enumerate}

We believe that the prescribed algorithms can be useful for the fast time-domain matched filtering calculation in the $\GstLAL$ pipeline. However, it is also necessary to investigate the practical challenges of implementing these algorithms in the current pipeline. In the current $\GstLAL$ pipeline, $\streamer$ multi-media framework has been used and $\HTC$ based distributed setup is developed to run all the steps involved in this pipeline \cite{cannon2020gstlal}. In recent work, Gittens {\em et al.} \cite{gittens2016multi} showed the adaptability of $\RMF$ algorithm in a $\Apache$-$\SPARK$ set-up. $\SPARK$-optimized code with enhanced computation power, making our algorithms much faster. Such fast computation of basis is unlikely to be possible for the current $\SVD$-based $\GstLAL$ framework, which is computationally expensive. In the future, we plan to explore further the possibility of integrating $\SPARK$ based $\RMF$ with the $\Gstreamer$ based matched-filtering setup, which can reduce the computational complexity of the matched filtering cost for the $\CBC$ search.

\begin{acknowledgments}
AR would like to thank Chad Hanna and Sarah Caudill for useful discussions on the $\GstLAL$ pipeline. The authors would like to thank Dilip Krishswamy for useful discussions and feedback.  This work was carried out with the
generous funding available from DST’s grant no. T-150 under the ICPS special call. 

\end{acknowledgments}


\appendix
\label{Sec:appendix}
\section{Implementation of Random Projection}
\label{Subsec:RP_Imp}
Let us consider a template matrix $\mathbf{H}$ has $2N_{T}$ number of rows $\{\bm{h}_{1},\cdots, \bm{h}_{i} \, : i = 1, 2, \cdots, 2N_{T}\} \in \mathbb{R}^{N_{s}}$. and $\exists$ a linear function $\Phi:\mathbb{R}^{N_{s}} \rightarrow \mathbb{R}^{\ell}$ which transformed $\bm{h}_{i} \rightarrow \bm{h}'_{i} = \Phi(\bm{h}_{i})$ such a way that for any pair of row vectors $\bm{h}_{i}$, $\bm{h}_{j}$ follows the following inequality.
\begin{equation}
(1-\epsilon)\, d(\bm{h}_{i}, \bm{h}_{j}) \leqslant d(\Phi(\bm{h}_{i})-\Phi(\bm{h}_{j})) \leqslant (1 + \epsilon)\, d(\bm{h}_{i}, \bm{h}_{j}) \, ,
\label{Eq:JL-lemma}
\end{equation}
where $d(\bm{h}_{i}, \bm{h}_{j})$ represents the the Euclidean distance between two row vectors $\bm{h}_{i}$ and $\bm{h}_{j}$ of the template matrix $\mathbf{H}$. Also, $\Phi(\bm{h}_{i}) = \bm{h}_{i} \, \mathbf{\Omega} \in \mathbb{R}^{\ell}$ represents the $\RP$ of each of the row vectors in $\ell$-dimensional space. The inequality shown in Eq.\ref{Eq:JL-lemma} known as $\Johnson$-$\Lindenstrauss$ ($\JL$)lemma. The $\JL$ lemma describes that by projecting any set of vectors from their original dimensional space to a lower-dimensional space, preserving the pairwise distances between the vectors can be possible within a distortion factor. The distortion factor measures the difference between the distance in the original and projected feature space. That implies, if the distortion factor is minimal, then the projected space dimension can preserve the pairwise distance between the vectors optimally. Therefore, the number of projected (reduced) dimensions is directly proportional to the distortion factor. This transformation from original dimensional to lower dimension space can be done easily by taking the linear combination of the rows of the template matrix ($\mathbf{H}_{2N_{T} \times N_{s}}$) with the columns of the $\RP$ matrix $\mathbf{\Omega}_{N_{s} \times \ell}$ \text{i.e.}, $\bar{\mathbf{H}}_{2N_{T} \times \ell} = \sum_{i = 1}^{2N_{T}}\sum_{j = 1}^{\ell} {\bm{h}_{i}\bm{\Omega}_{j}} = \mathbf{H} \mathbf{\Omega}$ such that $\mathbb{E}\big[\mathbf{\Omega} \, \mathbf{\Omega}^{T} \big] = \mathbb{I}$. This kind of matrix can be generated using several way \cite{achlioptas2001}, \cite{dasgupta1999}, \cite{li2006}. One of the well-known approaches is to use standard Gaussian distribution to generate the rows of the matrix $\mathbf{\Omega}: \Omega_{ij} \in \mathcal{N}(0,1)$. 

\begin{figure}[htbp]
\centering   
\begin{tabular}{cc}
\includegraphics[scale = 0.5]{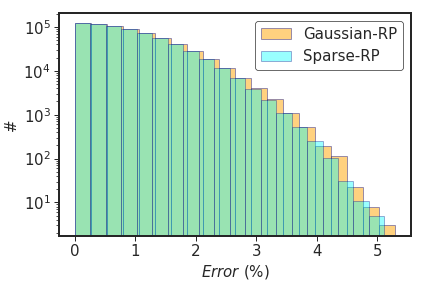}&
\includegraphics[scale = 0.5]{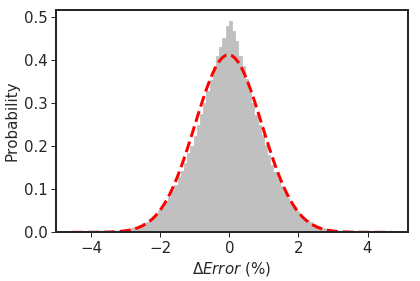}\\
(a) & (b)
\end{tabular}
\caption{(a) 
The figure shows the distribution of the error (percentile) of computing the pair-wise distance between a set of whitened template waveforms in the original dimensional space and projected space (with the dimension half of the original space) using the two $\RP$ operators, Gaussian and sparse. Here, the error is defined as $\Error = \frac{|d_{ij} - d'_{ij}| }{d_{ij}}$, where $d_{ij}$ is the distance between $i^{\texttt{th}}$ and $j^{\texttt{th}}$ rows of the template matrix $\mathbf{H}$ in the original space. Similarly, $d'_{ij}$ is the distance between $i^{\texttt{th}}$ and $j^{\texttt{th}}$ rows of the template matrix $\mathbf{H}$ in the projected space. The template matrix has been constructed by stacking a set of whitened template waveforms. (b) The difference between the percentile error distribution occurred due to two different $\RP$ operators. $\Delta Error$ defines the difference between the error obtained from two operators.
} 
\label{fig:RP-error}
\end{figure}
We have carried an experiment to demonstrate the inequality described by the $\JL$ lemma. 
The Figure-\ref{fig:RP-error} shows a histogram plot for the error of approximating the pair-wise distance for a set of whitened template waveform in the original as well in the projected (lower) dimensional space. 

\section{Subspace Iteration}
\label{Subsec:Sub_space_Iter}
\begin{figure}[htbp]
\centering{
\includegraphics[width=.50\textwidth]{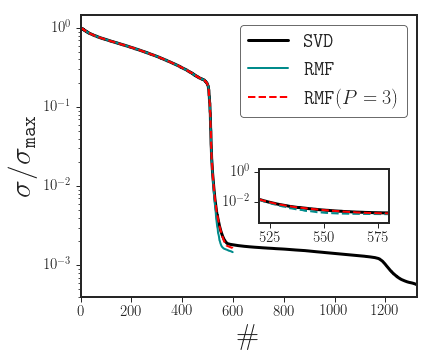}}
\caption{The figure shows a singular value spectrum of a template matrix $\mathbf{H} \equiv 1324 \times 225280$. The singular values are computed using $\SVD$, $\RMF$ with fixed rank-$\ell$, and $\RMF$ with power scheme. Here $\ell = 600$, and $P = 3$ is chosen.}
\label{Fig:Power-RMF}
\end{figure}

Alg-\ref{alg:RMF-Basic} describes the basic $\RMF$ of a template matrix $\mathbf{H}$. The accuracy of the basic $\RMF$ scheme depends on the characteristic of the profile of the singular values. If the singular values fall sharply, that implies that the number of important bases is less, whereas the profile is flat, implying that all bases have equal importance. Hence the error incurred due to the low-rank representation of the template matrix can be defined as a function of the singular values as follows: 
\begin{equation}
\epsilon = \|\mathbf{H} - \mathbf{H}^{(\ell)}\|_{F} = \Big( \sum_{i = \ell + 1}^{{\min \{2N_{T}, N_{s} \}}} {\sigma_i^2}\Big)^{\frac{1}{2}}
\label{Eq:Err_RMF}
\end{equation}
From the above relation (Eq.\ref{Eq:Err_RMF}), it is clear that if the singular spectrum of $\mathbf{H}$ decays slowly, $\RMF$ can have a high reconstruction error. In this scenario, Halko {\em et. al.} \cite{martinsson2011randomized} proposed to use power iteration scheme (For details see Alg-$4.4$ of \cite{halko2011} ). The main benefit of applying the power scheme is that the bases $\mathbf{U}$ and $\mathbf{V}$ remain the same after operating the power scheme on the template matrix $\mathbf{H}$. That implies, after applying the power method, the template matrix looks different, but the computed bases are the same as the original template matrix. Therefore the vectors of the range of the original template matrix $\mathbf{H}$ are still written as the linear combination of those basis vectors. However, the act of the power scheme on $\mathbf{H}$ changes the relative weights of the singular values and forcibly changes the spectrum of the singular values from slow decay to rapid decay. Formally, if the singular values of $\mathbf{H}$ are $\Sigma$, the singular values of $\big(\mathbf{H} \mathbf{H}^{T}\big)^{P} \mathbf{H}$ are $\Sigma^{2P+1}$. Here, we have outlined this relation mathematically. 
Let us consider, the template matrix $\mathbf{H}$ decomposes as follows: \\
$\mathbf{H} = \mathbf{U} \mathbf{\Sigma} \mathbf{V}^{T}$. 
Then the representation of the template matrix after applying the power method can be decomposed as follows:
\begin{equation}
\begin{split}
\Big(\mathbf{H} \mathbf{H}^{T}\Big)^{P} \, \mathbf{H} &=  \Big(\big(\mathbf{U} \, \mathbf{\Sigma} \, \mathbf{V}^{T} \big) \, \big(\mathbf{U} \, \mathbf{\Sigma} \, \mathbf{V}^{T} \big)^{T} \Big)^{P} \mathbf{H} \\
&= \Big(\mathbf{U \Sigma (V^{T} V) \Sigma U^{T}} \Big)^{P} \mathbf{H} \\
&= \Big(\mathbf{U} \mathbf{\Sigma}^{2} \mathbf{U}^{T} \Big)^{P} \mathbf{H} \ \big[\text{Since}, \, \mathbf{V}^{T} \mathbf{V} = \mathbb{I}_{2N_{T} \times 2N_{T}} \big]\\
&= \Big(\mathbf{U} \mathbf{\Sigma}^{2} \mathbf{U}^{T} \Big)^{P} \, \Big(\mathbf{U} \mathbf{\Sigma} \mathbf{V}^{T}\Big) \\
&= \mathbf{U} \, \mathbf{\Sigma}^{2P+1} \, \mathbf{V}^{T} \ \Big[\text{Using mathematical induction:} \, P = 1, 2, \cdots\Big]
\end{split}
\label{Eq:Sub-space}
\end{equation}
The above relation shown in Eq.\ref{Eq:Sub-space} holds due to the fact that the matrices $\mathbf{H}$ and $\big(\mathbf{H} \, \mathbf{H}^{T}\big)^{P}\, \mathbf{H}$ have the same left and right singular vectors. Since the $\SVD$ of $\Big( \mathbf{H} \mathbf{H}^{T} \Big)^{P} \mathbf{H}$ is $\mathbf{U} \mathbf{\Sigma}^{2P+1} \mathbf{V}^{T}$ and the obtained singular vectors obtained from the $\SVD$ is unique \cite{trefethen1997numerical}, hence, the singular vectors for $\mathbf{H}$ and $\big(\mathbf{H} \, \mathbf{H}^{T} \big)^{P}$ are the same. Hence it is clear from the above relation that the singular values spectrum decays exponentially with the number of power iterations, but the singular vectors remain same for both the matrices. 
From Eq.\ref{Eq:Err_RMF}, it is clear that if the singular values of the template matrix fall rapidly, then the reconstruction error is less. However, the error is large for a slowly decaying singular values spectrum. Hence, in practice, the minimal approximation error for  $\{ \sigma_{\ell + 1}, \cdots, \sigma_{\min \{2N_{T}, N_{s} \}} \}$ will be large. Therefore randomized scheme fails to provide the best $\ell$-dominant sub-space, and consequently, the scheme fails to provide a best rank-$\ell$ approximation of the template matrix. This problem can frequently occur for any large template matrix. To overcome this problem, one can sample $\big(\mathbf{H} \mathbf{H}^{T} \big)^{P} \mathbf{H}$ instead of sampling $\mathbf{H}$. One can show that $\big(\mathbf{H} \mathbf{H}^{T} \big)^{P} \mathbf{H}$ has the same left and right singular vectors as $\mathbf{H}$. However, the singular values of  $\big(\mathbf{H}\mathbf{H}^{T} \big)^{P} \mathbf{H}$ are the $(2P + 1)^{\th}$ power of the singular values of $\mathbf{H}$ , \text{i.e.}, $\sigma{\big(\mathbf{H} \mathbf{H}^{T} \big)^{P} \mathbf{H}} =  \sigma_i^{2P +1} \big(\mathbf{H} \big)$ [ As shown in Eq.\ref{Eq:Sub-space}]. This relation shows, even though the singular values spectrum of $\mathbf{H}$ falls slowly, then also the singular values of $\big(\mathbf{H} \mathbf{H}^{T} \big)^{P} \mathbf{H} \big)$ fall sharply. Hence it is always better to operate the $\RP$ on $\big(\mathbf{H} \mathbf{H}^{T} \big)^{P} \mathbf{H} \big)$ instead of $\mathbf{H}$. Finally, one has to compute the orthogonal basis vectors $\mathbf{Q}$ of $\big(\mathbf{H} \mathbf{H}^{T} \big)^{P} \mathbf{H} \mathbf{\Omega}$. Step-$2$ $\&$ $3$ of Alg-(\ref{alg:RMF-Basic}) can be replaced as follows:
\begin{itemize}
\item $\bar {\mathbf{H}} = \big(\mathbf{H} \mathbf{H}^{T} \big)^{P} \mathbf{H} \mathbf{\Omega}$
\item $\mathbf{Q} = qr \big( \bar {\mathbf{H}} \big) = qr \big( \big(\mathbf{H} \mathbf{H}^{T} \big)^{P} \mathbf{H} \mathbf{\Omega} \big) $
\end{itemize}
The optimal value of $P$ entirely depends on the low-rank structure of the template matrix. For a large template matrix, it is better to apply the power scheme to reduce the approximation error—however, the computational complexity increases due to the implementation of the power scheme. Hence, using the power scheme for a large template matrix in a distributed set-up is always recommended. Also, the power scheme can be easily adaptable in the distributed system architecture. In general, a small value of $P \, ( \, = 1, 2, 3)$ can also provide sufficient decay on the singular value spectrum; therefore, highly accurate results can be attained using more computational resources. 
The Figure-\ref{Fig:Power-RMF} demonstrates an example in which singular values spectrum using $\SVD$, $\RMF$ (Alg-\ref{alg:RMF-Basic}) and $\RMF$ with power iteration. The last few singular values obtained using $\RMF$ are not matched exactly with the original one. After applying the same algorithm with power iteration, we obtained the same top-$600$ singular values obtained from $\SVD$.

\section{Details of the random matrix factorization (\RMF)}
\label{Subsec:RMF-details}
\subsection{Fixed-rank \RMF \ (Category-\text{I})}
In this section, we have described two categories of fixed-rank $\RMF$ schemes. The fixed-rank $\RMF$ (Category-\text{I}) describes by Alg-\ref{alg:RMF-Basic}, whereas Alg-\ref{alg:RMF-Basic-II} represents the fixed-rank $\RMF$ (Category-\text{II}). The only difference between these two categories is the $\RP$ on the template matrix. In the first category, the row vectors of the template matrix are projected in the lower-dimensional space. However, in the latter category, the column vectors of the template matrix are cast in the lower dimensional space. That implies that the obtained set of basis vectors from the first category span the column space, and the obtained basis vectors from the second category span the row-space considering all the template waveforms represent a vector space. The mathematical concept is similar to the $\SVD$ decomposition, where $\mathbf{U}$ and $\mathbf{V}$ are the set of basis vectors of the row space and column space, respectively. The pictorial description of Alg-\ref{alg:RMF-Basic} is shown in Figure-\ref{Fig:RMF-Basic}.
\vspace{0.1\baselineskip}
\begin{algorithm}[H]
\DontPrintSemicolon
\KwIn{ Template Matrix \{$\mathbf{H}_{2N_{T} \times N_{s}} \in \mathbb{R}^{N_{s}}: 2N_{T} \geq N_{s} $\}.}
\KwOut{$\mathbf{Q}_{2N_{T} \times \ell}, \mathbf{B}_{\ell \times N_{s}}$}
$\mathbf{\Omega}_{N_{s} \times \ell } : \mathbf{\Omega}_{ij} \in \mathcal{N}(0,1)\} \; \&$  \atcp{Generate a random matrix}
$\bar{\mathbf{H}} = \mathbf{H \Omega} $ \atcp{Compression of the column space}
$\mathbf{Q} = qr(\bar{\mathbf{H}})$ \atcp{ Set of basis vectors of the projected space}
$\mathbf{B} = \mathbf{Q}^{T} \mathbf{H}$ \atcp{Surrogate matrix s.t. $ {\|{\mathbf{H}}\|_{F}}^{2} - {\|{\bf{B}}\|_{F}}^{2} \leq \delta$}
\caption{$\RMF$ with a fixed rank ($\ell$)}
\label{alg:RMF-Basic}
\end{algorithm}
\subsection{Fixed-rank \RMF \ (Category-\text{II})}
\vspace{0.5\baselineskip}
\begin{algorithm}[H]
\DontPrintSemicolon
\SetAlgoLined
\KwIn{Template Matrix \{$\mathbf{H} \in \mathbb{R}^{2N_{T} \times N_{s}}: 2N_{T} \leq N_{s} $\}, $\ell$}
\KwOut{$\mathbf{Q}_{2N_{T} \times \ell }$, $\mathbf{B}_{\ell \times N_{s}}$}
$\mathbf{\Omega}_{i} \in \mathbb{R} ^{N_{s} \times \ell}  \; : \mathbf{\Omega}_{ij} \; \in \mathcal{N}(0,1) \;$  \;
\For{P = 1, 2}{
${\bar{\mathbf{H}}}_{N_{s} \times \ell} = {\mathbf{H}^{T}}_{N_{s} \times 2N_{T}} \, \big(\mathbf{H}_{2N_{T} \times N_{s}} \, {\mathbf{\Omega}_{i}}_{N_{s} \times \ell} \big)$ \;
${\mathbf{\Omega_{i}}}_{N_{s} \times \ell} = \textrm{qr}(\bar{\mathbf{H}})$ \;
}
$\mathbf{Q_{i}}_{N_{s} \times \ell} = \mathbf{\Omega_{i}}$ \atcp{Consider top-$b$ basis vectors out of $\ell$ basis vectors.}
$\mathbf{B_{i}}_{2N_{T} \times \ell} = \mathbf{H}\,\mathbf{Q_{i}}$ \;

\caption{$\RMF$ ($\mathbf{H} \approx \mathbf{B} \, \mathbf{Q}^{T}$) with power scheme}
\label{alg:RMF-Basic-II}
\end{algorithm}
\vspace{0.5\baselineskip}
\subsection{Mathematical preliminaries of block-wise \RMF}
\textbf{Mathematical explanation of Null-space projection}: \\

\begin{figure}[htbp]
\centering{
\includegraphics[width = .65\textwidth]{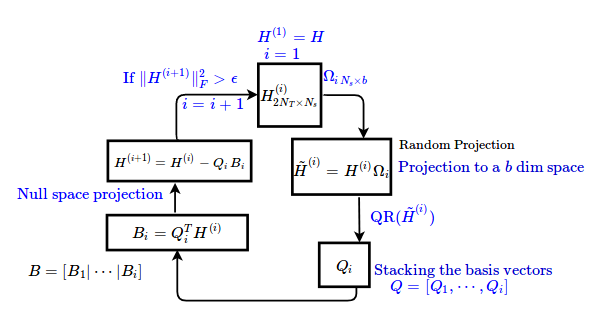}
\caption{This figure demonstrates an iterative method to compute basis vectors at a predefined average $\SNR$ loss. The first step is to project all the template waveforms in a small ($b$) dimensional space and then compute the top-$b$ basis vectors. If these set of basis vectors are sufficient to approximate the $\avgSNRLoss$ with a predefined error $\epsilon$, then no need to process further. The process has to be continued unless the error reaches the desired accuracy. The corresponding set of basis vectors are the final optimal set of bases.
 }
\label{Fig:BRMF}
}
\end{figure}

In this sub-section, we described the mathematical understanding of the null-space projection of the row vectors of the template matrix $\mathbf{H}$ which is an essential step for the $\RMF$ algorithm with a fixed error. Let us consider, a template matrix $\mathbf{H}_{2N_{T} \times N_{s}}$ decomposed as $\mathbf{H} \approx \mathbf{Q}_{2N_{T} \times \min{(2N_{T}, N_{s})}} \, \mathbf{B}_{\min(2N_{T}, N_{s}) \times N_{s}}$. Where $\mathbf{Q} = \{\bm{q}_{i}\}_{i = 1}^{\min\big(2N_{T},N_{s}\big)}$ as a set of basis and $\mathbf{B} = \{\bm{b}_{i}\}_{i = 1}^{\min \big(2N_{T},N_{s}\big)}$ as a set of coefficient vectors.  
\begin{enumerate}
\item The template matrix $\mathbf{H}_{2N_{T} \times N_{s}}$ can be decomposed as a sum of $\rank$-one matrices as follows:
\begin{equation}
\mathbf{H} = \mathbf{H}_{1} + \mathbf{H}_{2} + \cdots + \mathbf{H}_{\min\big(2N_{T}, N_{s}\big)} \, ,
\label{rankOne_Decom}
\end{equation}
where $\mathbf{H}_{i} \Big( = \bm{q}_{i} \bm{b}_{i}, i = 1,2,\cdots,\min(2N_{T}, N_{s})\Big)$ represents the $\rank$-one matrices. 
\item Using the concept of $\rank$-one matrix decomposition, one can similarly decompose a matrix $\mathbf{H}$ into a sum of $\rank$-$b$ matrix as follows:
\begin{equation}
\mathbf{H} = \mathbf{Q}_{1} \mathbf{B}_{1} + \mathbf{Q}_{2} \mathbf{B}_{2} + \cdots + \mathbf{Q}_{\min \big(2N_{T}, N_{s}\big)} \mathbf{B}_{\min \big(2N_{T}, N_{s} \big)} \, ,
\label{rank-b_Decom}
\end{equation}
where the dimension of each basis matrix $\mathbf{Q}_{i}: i = 1, 2, \cdots, \min(2N_{T}, N_{s})$ is $b \times \min(2N_{T}, N_{s})$, and dimension for the coefficient matrix $B_{i}: i = 1, 2, \cdots, \min(2N_{T}, N_{s})$ is $\min(2N_{T}, N_{s}) \times N_{s}$. For simplicity, we considered that all matrices $\mathbf{Q}_{i}$ contains same number of basis \text{i.e.}, $b$. 
\item Let us define a $\residual$ matrix $\mathbf{R}$ as follows $\mathbf{R} = \mathbf{H} - \mathbf{Q}\mathbf{B}$. \\
After first iteration, it can be computed as: $\mathbf{R}^{(1)} = \mathbf{H} - \mathbf{Q}_{1} \mathbf{B}_{1}$. 
If the $\mathbf{Q}_{1}\mathbf{B}_{1}$ represents the $b$-rank approximation of the template matrix $\mathbf{H}$, then the residual matrix (after the first iteration) implies that it only contains the contribution of the $\Big(\min(2N_{T}, N_{s})-b\Big)$ number of the basis vectors as the contribution for the top-$b$ basis vectors already subtracted out from it. \\
In the second iteration, we want to compute the second set of important basis vectors $b+1, b+2, \cdots, b+b$. Therefore, to compute this set, we can use the residual matrix $\mathbf{R}^{(1)}$ as it contains a contribution of the rest of the basis vectors. Thus, we need to apply the $\QR$ decomposition of $b$-dimensional representation of the residual matrix $\mathbf{R}^{(1)}$ to obtain the set of basis $\mathbf{Q}_{2}$ \text{i.e.} $\mathbf{Q}_{2} = \texttt{qr}\big({\mathbf{R}^{(1)} \mathbf{\Omega_{2}}}\big)$. Further, the coefficient matrix $\mathbf{B}_{2}$ can be computed by projecting all the basis vectors on $\mathbf{R}^{(1)}$, \text{i.e.} $\mathbf{B}_{2} = \mathbf{Q}_{2}^{T} \mathbf{R}^{(1)}$. This process is called the null-space projection of the basis vectors. In every iteration, we subtract the energy (contribution) of a specific set of bases, and then the next set of importance bases is computed from the residual matrix. In this way, we reach the null-space after a $i^{\texttt{th}}$ iteration, and the total number of basis vectors up to that iteration is called the rank of the matrix, \text{e.g.}, if in each iteration, we get top-$b$ basis vectors and the rank of the matrix is $\ell$, then after $i^{\texttt{th}}$ iteration we should obtain top $\ell$ basis vectors, that implies $\ell = ib$. The rest of $(\ell-\min(2N_{T}, N_{s})$ basis vectors are less important and after $i^{\texttt{th}}$ iteration the energy of the matrix is also minimal, therefore, these $(\ell-\min(2N_{T}, N_{s})$ are considered the basis vectors for the null-space. We can compute the $\residual$ matrix after the second iteration as follows:
$\mathbf{R}^{(2)} = \mathbf{H} - \mathbf{Q}_{2}\mathbf{B}_{2} = \mathbf{H} - \mathbf{Q}_{2} \Big(\mathbf{Q}_{2}^T \mathbf{R}^{(1)}\Big)$. 
This implies, after second iteration, we already got the $2b$ of basis vectors combining $\big[\mathbf{Q}_{1}, \mathbf{Q}_{2}\big]$, and the residual matrix now contains the energy of the space span by the maximum $(\min(2N_{T}, N_{s})-2b)$ basis vectors. 
Similarly, we can generalize the calculation of the residual part after $i^{\texttt{th}}$ iteration as follows:
$\mathbf{R}^{(i)} = \mathbf{H} - \mathbf{Q}_{i}\mathbf{B}_{i} = \mathbf{H} - \mathbf{Q}_{i} \Big(\mathbf{Q}_{i}^{T} \mathbf{R}^{(i-1)}\Big)$. \\
\end{enumerate}
We used the null-space projection for the computation of the  basis and coefficient matrices $\{\mathbf{Q_{i}}\}_{i=1}^{r}$ and $\{\mathbf{B_{i}}\}_{i=1}^{r}$ in each iteration in the block-wise $\RMF$ algorithm. 
We first initiate the algorithm by setting $\mathbf{H}^{(0)} = \mathbf{H}$. Then we follow following step for each iteration $i = 1, 2, \cdots \ell$.
\begin{enumerate}
\item Computation of the basis vectors from a set of vectors from a lower-dimensional row-space. $\mathbf{Q}_{i} = \texttt{orth} \big(\mathbf{H}^{(i-1)} \mathbf{\Omega}_{i} \big)$
In the first iteration, the row vectors of the data matrix are projected to a $b$-dimensional space using $\RP$ matrix ${\mathbf{\Omega}_{i}}_{b \times N_{s}}$. However, from the first iteration onwards, we computed the residual energy (in terms of Frobenius norm) of the data matrix, and from the second iteration onwards, the set of basis, \text{e.g.}, $\{\mathbf{Q}_{2}, \mathbf{Q}_{3}, \cdots$ has been computed using the null-space projection of the row-space obtained from the previous iteration.
\begin{align*}
\mathbf{Q}_{2} &= \texttt{qr} \big(\mathbf{R}^{(1)} \mathbf{\Omega}_{2} \big) \\
&= \texttt{qr} \big(\mathbf{H} \mathbf{\Omega}_{2} - \mathbf{Q}_{1} \mathbf{B}_{1} \mathbf{\Omega}_{2} \big)
\end{align*}
\item In each iteration, it is required to obtain the coefficient matrix to measure the corresponding error. $\mathbf{B}_{i} = {\mathbf{Q}_{i}}^{T} \mathbf{H}^{(i-1)}$. From the second iteration onwards, $\mathbf{H}^{(i-1)}$ represents the null-space projection. Therefore, the coefficient matrix is computed using the residual matrix of the previous iteration. That implies that the coefficient vectors are computed by projecting the $\residual$ matrix onto the block of basis vectors. However, using the following steps, one can show mathematically that this is exactly equivalent to the projection of the orthogonal basis onto the original data matrix if the computed basis vectors in the current stage are orthogonal to the previous stage, \text{i.e.}, $\mathbf{Q}_{i-1}^{T} \,\mathbf{Q}_{i} = 0$.
\begin{align*}
\mathbf{B}_{2}
&= \mathbf{Q}_{2}^T \mathbf{R}^{(1)} \\
&= \mathbf{Q}_{2}^T \mathbf{H}- \mathbf{Q}_{2}^T \mathbf{Q}_{1} \mathbf{B}_{1} \\
&= \mathbf{Q}_{2}^T \mathbf{H} \, \big[ \mathbf{Q}_{j}^T \mathbf{Q}_{i} = 0, i \neq j \big]
\end{align*}
\item In each iteration, it is required to trace the residual energy of the data matrix, and it can be done by subtracting out the $b$-rank approximation of the matrix from the data matrix in each iteration.
$\mathbf{H}^{(i)} = \mathbf{H}^{(i-1)} - \mathbf{Q}_{i} \mathbf{B}_{i}$ 

$\mathbf{H}^{(i)}$ holds precisely the residual remaining after $i^{\texttt{th}}$ step.
This means that incorporating adaptive rank determining can be possible by simply computing the residual energy in each iteration \text{i.e.} $\| \mathbf{H}^{(i)} \|_{F} \leq \epsilon$. The pictorial representation of this block-wise $\RMF$ has been demonstrated in Fig.\ref{Fig:BRMF}. 
\end{enumerate}
\subsection{Type-\text{I} \ \RMF\ (row-wise)}
Alg-\ref{alg:DrMF_r} shows the computation of a standard set of basis vectors by combining all the sub-banks of a specific template bank. We assumed that the number of essential basis vectors by combining all the sub-banks is $\ell$. Hence, the algorithm provides top-$\ell$ basis vectors and their corresponding coefficients. Further, we assumed that the whole template matrix is not be stored in a single machine due to the size of the template matrix. Hence, the entire template matrix is divided into a sub-template matrix. These matrices are placed over a set of local machines. These sub-template matrices are similar to the sub-banks defined in the $\GstLAL$ pipeline. Currently, for the non-precession waveforms, the sub-bank division in the $\GstLAL$ has been done based on chirp-mass ($\mathcal{M}$) and duration of the waveforms. We have also followed the same way of obtaining sub-banks. In $\GstLAL$, each sub-banks are treated separately, and computed basis vectors span the waveforms from those sub-banks only. Thus the basis vectors obtained from each sub-banks are independent of each other. Here we present a scheme ( described by Alg-\ref{alg:DrMF_r}) to get a basis combining all the sub-banks. That implies that the obtained basis vectors are common to all the sub-banks. 

In this section, we have described Alg-\ref{alg:DrMF_r} and estimates the required storage memory in each step. Let us consider, a template bank has $N_{T}$ number of template waveform and each of the waveform has $N_{s}$ time samples. Hence, the size of the template matrix is $H_{2N_{T} \times N_{s}}$. Also, consider that we have $n_{b}$ number of individual machines and each machine has $16 \GB$ \, $\RAM$. Each machine can contains $N_{T_{b}}$ number of templates. Hence, the total number of templates in a template bank is $N_{T} = n_{b} \, N_{T_{b}}$. Let $\ell$ is the required number of important bases corresponding to a specific $\avgSNRLoss$. 
In $\RMF$, the template waveforms are projected in a lower-dimensional space using a $\RP$ operator. In Alg-\ref{alg:DrMF_r}, the compressed version (or the projected version) of the waveforms for each sub-banks is formulated using a $\RP$ matrix $\mathbf{\Omega}$; the step is shown in the step-$2$. All the sub-banks are distributed over different machines; hence, we have two options to perform the projection operation. The first option is that we can provide the same $\mathbf{\Omega}$ to all machines and then complete the projection operation on each sub-bank separately. The alternative option is that we can generate the same $\Omega$ for all machines. The first option is computationally expensive, as there is a communication cost is involved for the passing of the same $\mathbf{\Omega}$ to all machines. However, the second option does not require any communication cost. Hence it is optimal. But the problem is that, in general, each machine can generate different $\mathbf{\Omega}$, which potentially affects the projection procedure as the waveforms from all the sub-banks should be projected using the same $\RP$ vectors to the lower-dimensional subspace. It can be resolved if we use a fixed seed for the generation of all random vectors of $\mathbf{\Omega}$ for all the machines. 
The step $3, 4$ of Alg-\ref{alg:DrMF_r} needs to be done in a central node connected with all the other local machines. This node can be called a "master node." Step- $3$ shows the collection of all lower-dimensional represents of all template waveforms from all the sub-banks. In step-$4$, the basis is computed from that $\ell$ dimensional sub-space. 
To estimate each step's required storage memory, let us consider a template matrix containing $10^{5}$ number of template waveform, and each waveform has $10^{6}$ of sample points. The storing of such a large waveform required $160 \GB$ disk space. Let us consider; each machine has $16 \GB$ $\RAM$ memory space. Therefore, each machine can contain $10^{2}$ number of template waveform that means the size of the each sub-matrix $\mathbf{H_{i}}$ is $2N_{T_{b}} \times N_{s} \equiv 200 \times 10^{6}$. The required memory to store such a sub-bank is $1.6 \GB$. Let us consider $\ell = 10^{4}$, therefore to generate the $\RP$ matrix required memory space is $80 \GB$. Since the machine does not have that much memory, it is impossible to generate the whole $\RP$ matrix in one machine. We can generate a block of the $\RP$ matrix at a time. Therefore $\mathbf{\Omega}$ defined in step-$1$ can be generated as follows:
$\mathbf{\Omega} = \Big[ {\mathbf{\Omega}_{1}}_{2N_{T_b} \times b} | {\mathbf{\Omega}_{2}}_{2N_{T_b} \times b} | \cdots | {\mathbf{\Omega}_{j}}_{2N_{T_b} \times b} \Big]$. 
If we consider $b = 10^{3}$ and $j = 10$, then $\mathbf{\Omega}_{j}$ can be of size $10^{6} \times 10^{3}$. The required memory to generate each of $\RP$ block is $1$ $\GB$. Hence, the step-$2$ can be computed using $\RP$ block as follows:
$\bar{\mathbf{H}}_{i} = \mathbf{H}_{i} \, \mathbf{\Omega}_{1} + \mathbf{H}_{i} \, \mathbf{\Omega}_{2} + \cdots + \mathbf{H}_{i} \, \mathbf{\Omega}_{j}$. 
To store each of these sub-matrices $\mathbf{H}_{i} \, \mathbf{\Omega}_{j}$ required memory is $16 \MB$. Thus, the required memory to store all these sub-matrices and the final output matrix $\bar{\mathbf{H}}_{i}$ shown in step-$2$ is $320 \, ( = 16 \times 10 + 160 ) \MB$. The required memory to store randomly projected template matrix $\bar{\mathbf{H}}$ shown in step-$3$ is $16 \GB$. Note that this matrix is stored in the master node. The master node should have a large memory space compared to the local distributed nodes (machines). There is a communication cost also involved in this step. $2 \times 10^{6}$ number of floating-point operations are required to send the matrix $\bar{\mathbf{H}}_{i}$ from the individual machine to the master node. Overall, the master node required $32 \GB$ of memory space to save $\bar {\mathbf {H}}$ and $\mathbf{Q}$ in the memory. Again there is also a $10^{6}$ number of floating-point operations are needed to send back again the corresponding bases $\mathbf{Q}_{i}$ to the individual machine from the master node. To perform the $\QR$ decomposition in step-$4$ required $2 \times 10^{13}$ floating-point operations. The required memory to store $\mathbf{Q}_{i}$ in each machine is $16 \MB$. Also, to store $B_{i}$ it is required $10 \GB$. Another communication involves from local machines to the central node to compute the computation of the coefficient matrix $\mathbf{B}$ for which the communication cost is $10^{10}$ number of floating-point operations. Alg-\ref{alg:BDrMF_r2} describes the block-$\RMF$ version of Alg-\ref{alg:DrMF_r} at a fixed error. However, if we combine $\RMF$ with $\MP$, then the computation of the coefficient matrix $\mathbf{B}$ is not required, and in that case, we should not count the estimated cost for the construction of the coefficient matrix. 
\vspace{0.5\baselineskip}
\begin{algorithm}[H]
\DontPrintSemicolon
\KwIn{Sub-bank \{$(\mathbf{H}_{i})_{2{N_{T}}_{b} \times N_{s}} $\}}
\KwOut{$\mathbf{Q}_{2N_{T} \times \ell }$, $\mathbf{B}_{\ell \times N_{s}}$}
$\mathbf{\Omega} \in \mathbb{R}^{N_{s} \times \ell}: \Omega_{ij} \in \mathcal{N}(0,1)$\\
$({\bar{\mathbf{H}}_{i}})_{2{N_{T}}_{b} \times \ell} = \mathbf{H}_{i} \, \mathbf{\Omega}$\\
$(\bar{\mathbf{H}})_{2N_{T} \times \ell} = [\mathbf{\bar{H}}_{1}|\mathbf{\bar{H}}_{2}|\cdots|\mathbf{\bar{H}}_{i}]$ \atcp{Row-wise stacking}
$\mathbf{Q}_{2N_{T} \times \ell}, \mathbf{R}_{\ell \times \ell} = \text{qr}(\bar{\mathbf{H}})$\\
$(\mathbf{Q}_{i})_{2{N_{T}}_{b} \times \ell} = \mathbf{Q}[(i-1){N_{T}}_{b} \, : i{N_{T}}_{b}, :]; i = 1, 2, \cdots, n_{b}$\\
${\mathbf{B}_{i}}_{_{\ell \times N_{s}}} = {\mathbf{Q}_{i}}^{T} \, \mathbf{H}_{i}$\\
$\mathbf{B}_{\ell \times N_{s}} = \sum_{i = 1}^{n_{b}}{\mathbf{B}_{i}}$\\
 
\caption{Type-I $\RMF$ (row-wise)}
\label{alg:DrMF_r}
\end{algorithm}

\vspace{0.5\baselineskip}
\begin{algorithm}[H]
\DontPrintSemicolon
\KwIn{Sub-bank \{$\big(\mathbf{H}_{i}\big)_{2{{N_{T}}_{b}} \times N_{s}} $\}}
\KwOut{$\mathbf{Q}_{2N_{T} \times \ell }$, $\mathbf{B}_{\ell \times N_{s}}$}
\For{i = 1,2,3, $\cdots$}{
$\mathbf{\Omega}_{i} \in \mathbb{R}^{N_{s} \times b}: \Omega_{ij} \in \mathcal{N}\big(0, 1 \big)$\\
$({\bar{\mathbf{H}}_{i}})_{2{{N_T}_{b}} \times b} = \mathbf{H}_{i} \, \mathbf{\Omega}_{i}$\\
$(\bar{\mathbf{H}})_{2N_{T} \times b} = \big[ \mathbf{\bar{H}}_{1}|\mathbf{\bar{H}}_{2}|\cdots|\mathbf{\bar{H}}_{i} \big]$ \atcp{Row-wise stacking}
$\mathbf{Q}_{2N_{T} \times b}, \mathbf{R}_{b \times b} = \text{qr}(\bar{\mathbf{H}})$\\
$\mathbf{Q}_{i} = \textrm{qr} \, \big (\mathbf{Q}_{i} - \sum_{j = 1}^{i-1}{\mathbf{Q}_{j} {\mathbf{Q}_{j}}^{T} \mathbf{Q}_{i}}\big )$ \\
$\big(\mathbf{Q}_{i}\big)_{2{{N_{T}}_{b}} \times b} = \mathbf{Q}\big[(i-1){N_{T}}_{b} \, : i{N_{T}}_{b}, : \big]; i = 1, 2, \cdots, n_{b}$ \\
${\mathbf{B}_{i}}_{{b \times N_{s}}} = {\mathbf{Q}_i}^{T} \, \mathbf{H}_{i}$\\
$\mathbf{B}_{b \times N_{s}} = \sum_{i = 1}^{n_{b}}{\mathbf{B}_{i}}$\\
$\mathbf{H} = \mathbf{H} - \mathbf{Q}_{i} \, \mathbf{B}_{i}$ \\ 
\If{$\|\mathbf{H}\|_{F} \geq \langle \frac{\delta \rho}{\rho}\rangle$}{$\mathbf{Q} = \big[\mathbf{Q}_{1}|\mathbf{Q}_{2}|\mathbf{Q}_{3}|\cdots|\mathbf{Q}_{i}\big]$ \\
$\mathbf{B} = \big[{\mathbf{B}_{1}}^{T}|{\mathbf{B}_2}^{T}|\cdots|{\mathbf{B}_i}^{T}\big]$}}
\caption{Type-$\text{I}$ $\RMF$ (row-wise) at a fixed error)}
\label{alg:BDrMF_r2}
\end{algorithm}

\subsection{Type-\text{II} \ \RMF \ Algorithm (column-wise division)}
The Alg-\ref{alg:DrMF_c} shows the scheme of obtaining basis at a fixed rank-$\ell$ over a distributed architecture framework. In this case, we assumed that the number of rows of a template matrix is less than the number of columns. Hence, the columns of the template waveforms are distributed over a set of local machines.

The prescribed framework is similar to the split-bank framework of the $\GstLAL$ pipeline. In a split-bank (or time-slices) framework, the template waveforms are sliced over time for a specific sub-bank. In $\GstLAL$, split banks are made by considering different sampling frequencies for the different parts of the waveform. Individual $\SVD$ has performed to a specific region to obtain the basis vectors. This approach helps handle a long waveform. We split a template into several regions for our purpose, but the sampling rate for generating these parts is the same. Also, we have combined all the parts to obtain a set of bases by combining all the split-banks. Here we assumed that each local machine could store only a fraction of the sampling points.

\begin{algorithm}[H]
\DontPrintSemicolon
\KwIn{Split bank \{$\mathbf{H}_{2N_{T} \times N_{s}} = \big[\mathbf{H}_{1}|\mathbf{H}_{2}|\cdots|\mathbf{H}_{i}\big] :\mathbf{H}_{i} \in \mathbb{R}^{2N_{T} \times (N_{s}/p)}: 2N_{T} \leq (N_{s}/p) $\}.}
\KwOut{$\mathbf{Q}_{2N_{T} \times \ell }$, $\mathbf{B}_{\ell \times N_{s}}$}
$\mathbf{\Omega} \in \mathbb{R}^{(N_{s}) \times \ell}: \Omega_{ij} \in \mathcal{N}(0,1)$\\
$\mathbf{\Omega}_{i} = \mathbf{\Omega}[(i-1)(N_{s}/p):i(N_{s}/p), :], i = 1, 2, \cdots, p$\\
$({\bar{\mathbf{H}}_{i}})_{2{N_{T}} \times \ell} = \mathbf{H}_{i} \, \mathbf{\Omega}_{i}$\\
$(\bar{\mathbf{H}})_{2N_{T} \times \ell} = \sum_{i = 1}^{n_{b}}{\bar{\mathbf{H}}_{i}}$\\
$\mathbf{Q}_{2N_{T} \times \ell}, \mathbf{R}_{\ell \times \ell} = \text{qr}(\bar{\mathbf{H}})$\\
${\mathbf{B}_{i}}_{_{\ell \times (N_{s}/p)}} = \mathbf{Q}^{T} \, \mathbf{H}_{i}$\\
$\mathbf{B}_{\ell \times N_{s}} = [\mathbf{B}_{1}|\mathbf{B}_{2}|\cdots|\mathbf{B}_{i}]$ \atcp{Column-wise stacking}
\caption{Type-II $\RMF$ (column-wise)}
\label{alg:DrMF_c}
\end{algorithm}
\begin{algorithm}[H]
\DontPrintSemicolon
\KwIn{Split bank \{$\mathbf{H}_{2N_{T} \times N_{s}} = \big[\mathbf{H}_{1}|\mathbf{H}_{2}|\cdots|\mathbf{H}_{i}\big] : \mathbf{H}_{i} \in \mathbb{R}^{2N_{T} \times \big(N_{s}/p\big)}: 2N_{T} \leq \big(N_{s}/p\big) $\}.}
\KwOut{$\mathbf{Q}_{2N_{T} \times \ell }$,
$\mathbf{B}_{\ell \times N_{s}}$}
\For{i = 1,2,3, $\cdots$}{
$\mathbf{\Omega}_{i} \in \mathbb{R}^{(N_{s} \times b)}: \Omega_{ij} \in \mathcal{N}\big(0, 1 \big)$\\
${\mathbf{\Omega}_{i}}^{(i)} = \mathbf{\Omega}\big[(i-1)(N_{s}/p):i(N_{s}/p), :\big], i = 1, 2, \cdots, p$\\
$\big({\bar{\mathbf{H}}_{i}}\big)_{2 N_{T} \times b} = \mathbf{H}_{i} \, {\mathbf{\Omega}_{i}}^{(i)}$\\
$\big(\bar{\mathbf{H}}\big)_{2N_{T} \times b} = \sum_{i = 1}^{n_{b}}{\mathbf{H}_{i}}$\\
$\mathbf{Q}_{2N_{T} \times b}, \mathbf{R}_{b \times \ell} = \textrm{qr}(\bar{\mathbf{H}})$\\
${\mathbf{B}_{i}}_{{b \times (N_{s}/p)}} = \mathbf{Q}^{T} \, \mathbf{H}_{i}$\\
$\mathbf{B}_{b \times N_{s}} = \big[\mathbf{B}_{1}|\mathbf{B}_{2}|\cdots|\mathbf{B}_{i}\big]$ \atcp{Column-wise stacking}
$\mathbf{H} = \mathbf{H} - \mathbf{Q}_{i} \, \mathbf{B}_{i}$ \\ 
\If{$\|\mathbf{H}\|_{F} \geq \langle \frac{\delta
\rho}{\rho}\rangle$}{
$ \mathbf{Q} = \big[\mathbf{Q}_{1}|\mathbf{Q}_{2}|\mathbf{Q}_{3}|\cdots|\mathbf{Q}_{i}\big]$ \\
$\mathbf{B} = \big[{\mathbf{B}_{1}}^{T}|{\mathbf{B}_2}^{T}|\cdots|{\mathbf{B}_{i}}^{T}\big]$}
} 
\caption{Type-\text{II} $\RMF$ (column-wise) with a fixed error}
\label{alg:BDrMF_c2}
\end{algorithm}

\begin{algorithm}[H]
\DontPrintSemicolon
\SetAlgoLined
\KwIn{Template Matrix \{$\mathbf{H} \in \mathbb{R}^{2N_{T} \times N_{s}}: 2N_{T} \leq N_{s} $\}, $\avgSNRLoss$}
\KwOut{$\mathbf{Q}_{2N_{T} \times k }$, $\mathbf{\tilde{B}}_{k \times N_{s}}$}
\For{$i = 1,2,\cdots$}{
$\mathbf{\Omega}_{i} \in \mathbb{R}^{N_{s} \times b}  \; : \mathbf{\Omega}_{ij} \; \in \mathcal{N}\big(0, 1\big) \;$  \;
\For{P = 1, 2}{
${\bar{\mathbf{H}}}_{N_{s} \times b} = {\mathbf{H}^{T}}_{N_{s} \times 2N_{T}} \, \big(\mathbf{H}_{2N_{T} \times N_{s}} \, {\mathbf{\Omega}_{i}}_{N_{s} \times b} \big)$ \;
${\mathbf{\Omega_{i}}}_{N_{s} \times b} = \textrm{qr}(\bar{\mathbf{H}})$ \;
}
$\mathbf{Q}_{i} = \mathbf{\Omega}_{i}$ \atcp{Consider top-$b$ basis vectors out of $\ell$ basis vectors.}
$\mathbf{B}_{i} = \mathbf{H}\,\mathbf{Q}_{i}$ \;
$\mathbf{H} = \mathbf{H} - \bf{Q}_{i} \, \mathbf{B}_{i}$\;
\If{$\|\mathbf{H}\|_{F} \geq \avgSNRLoss$}{
$ \mathbf{Q} = \big[\mathbf{Q}_{1}|\mathbf{Q}_{2}|\mathbf{Q}_{3}|\cdots|\mathbf{Q}_{i} \big]$ 
}
}
$\text{Choose}\, k \, \text{based on}\, \ell. $ \;
$\mathbf{Q} = \mathbf{Q}\big[:, 1:k\big]$\;
\For{$\alpha$ = 1,2,3, $\cdots$, $2N_{T}$}{
$\text{Find} \, \vec{\tilde{b}}^{\alpha} : H^{\alpha} \approx \vec{\tilde{b}}^{\alpha}_{1 \times k} \, \mathbf{Q}_{k \times N_{s}} $ \atcp{Using Matching Pursuit algorithm}
${\mathbf{\tilde{B}}}_{2N_{T} \times k} = [\vec{\tilde{b}}^{1}|\vec{\tilde{b}}^{2}|\cdots|\vec{\tilde{b}}^{\alpha}]$ \atcp{Row-wise stacking}\;
}
\caption{\RMF with fixed error + sparse coefficient using \MP. }
\label{alg:RMF-MP-I}
\end{algorithm}

\section{Construction of a sparse reconstruction matrix (\RMF + \MP)}
Alg-\ref{alg:RMF-MP-I} describes the hybrid scheme for obtaining basis vectors along with the sparse coefficient by combining block-wise $\RMF$ and $\MP$ algorithm. A set of basis vectors is first obtained corresponding to a fixed error, which shows in the steps $1-13$ of the algorithm. Further, the first $k$ number of basis vectors out of $\ell$ are used to construct the sparse coefficient vectors with $k$ non-zero elements. The steps $14-20$ of the algorithm show this sparsification of coefficient vectors using the $\MP$ algorithm.   

\section{ Applicability of \RMF \ algorithms over the \CBC \ parameter space}
\label{Subsec:Prac_imp_RMF}
This section shows the pictorial representation of the applicability of the various proposed $\RMF$ schemes in the different regimes of the parameter space for the search of $\GW$ signal from $\CBC$ sources.  
\begin{figure}[]
\centering{
\includegraphics[width = .75\textwidth]{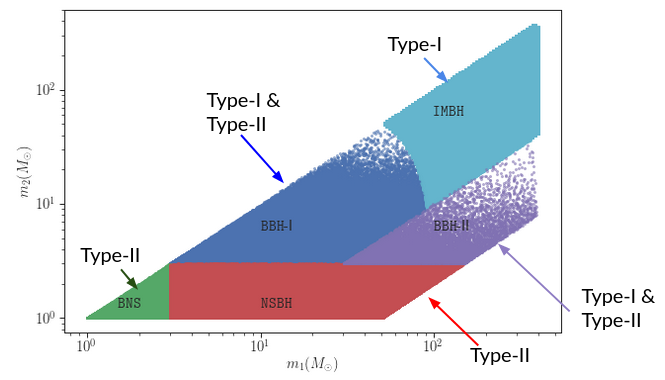}
\caption{The figure shows the applicability of prescribed $\RMF$ algorithms (Type-$\text{I}$, Type-$\text{II}$) for the different parameter space regimes. Figure-\ref{Fig:TD-Comp} shows the duration of the waveform for the various $\CBC$ sources with a fixed lower cut-off frequency. For $\BNS$ and $\NSBH$ systems, the waveform duration is much longer than $\BBH$ and $\IMBH$ systems. Hence, Type-$\text{I}$ $\RMF$ can be applicable for $\IMBH$ in which the waveforms are shorter than other systems. Similarly, Type-$\text{II}$ $\RMF$ is directly relevant to $\BNS$ and $\NSBH$ systems. However, for the $\BBH$ system, both Type-$\text{I}$ and Type-$\text{II}$ schemes are applicable.
}
\label{Fig:Parameter-space}
}
\end{figure}
\begin{figure}[]
\centering{
\includegraphics[width = .80\textwidth]{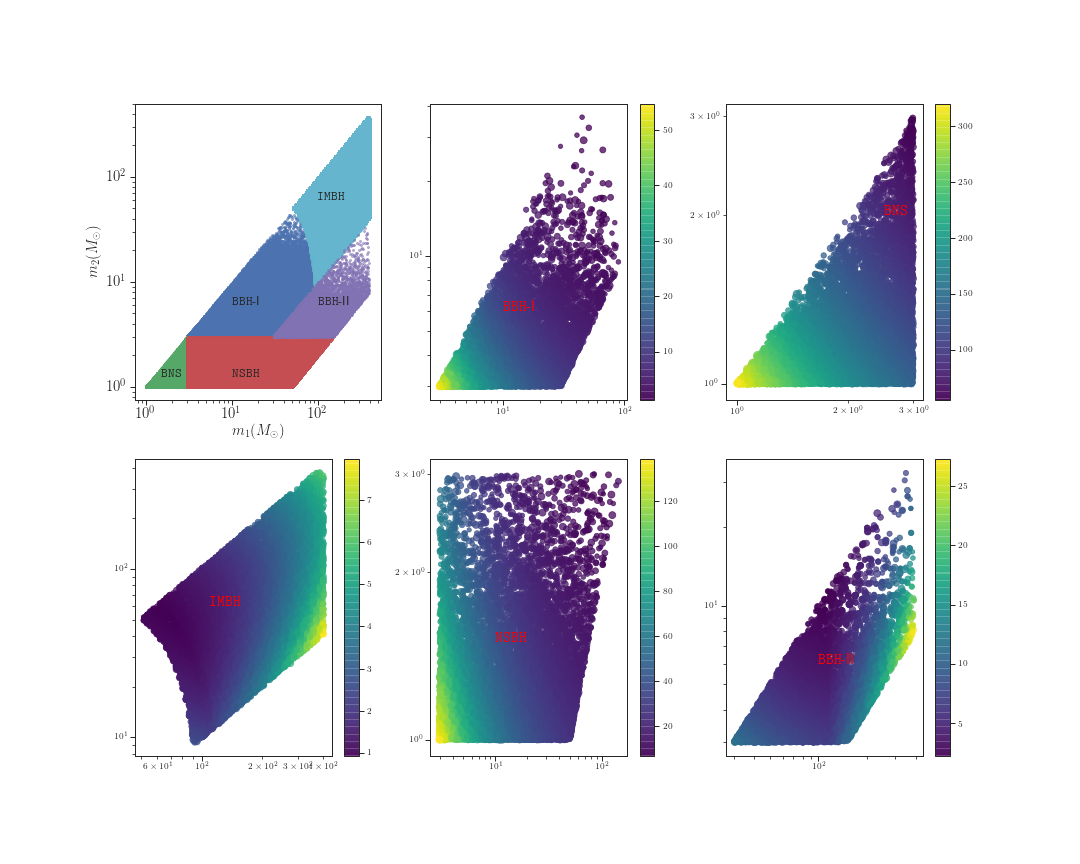}
\caption{The figure shows a two-dimensional template bank on the mass components $m_{1, 2}$. The color bar indicates $10^{4}$ randomly picked template waveforms' time-duration in the specific parameter regime, including $\BNS$, $\BBH$, $\NSBH$, and $\IMBH$. The template bank boundary has been chosen as same as taken as $\GstLAL$ pipeline in the third observing runs ($\O 3$) \cite{abbott2020gwtc}. The details of the parameter space are as follows. 
(a) $\BNS$ region: $1$-$3$ in solar masses, $15$-$1024$ $\Hz$, aligned or anti-aligned spins to $0.05$, match = $.99$.
(b) $\BBH$-low-$q$: $3$-$100$ in solar masses, $15 - 1024$ $\Hz$, aligned or anti-aligned spins to $0.999$, match = $.99$, mass ratios under $10$.
(c) $\NSBH$: $1$-$3$ and $3$-$150$ in solar masses, $15$-$1024$ $\Hz$, aligned or anti-aligned spins to $0.05$ and $.999$, match = $.97$,  mass ratios under $50$.
(d) $\IMBH$: $10$-$400$ in solar masses, $10$-$1024$ $\Hz$, aligned or anti-aligned spins to $0.99$.
(e) Other $\BBH$: $3$-$100$ in solar masses, $15$-$1024$ $\Hz$, aligned or anti-aligned spins to $.999$, match = $0.97$, mass ratios = $10$-$33$.
}
\label{Fig:TD-Comp}
}
\end{figure}

\begin{figure}[]
\centering{
\includegraphics[width = 0.65\linewidth]{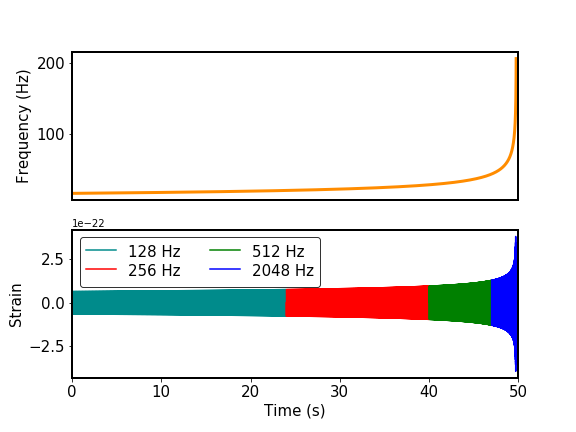}
\caption{The figure shows the evolution of a time-domain template waveform with increased frequency components. We have generated the waveform corresponding to a $\NSBH$ system, in which the component masses $m_{1} = 1.3 \, M_{\odot}$, $m_{2} = 20.0 \, M_{\odot}$, the lower cut-off frequency $15$ $\Hz$ are chosen. The $\texttt{TaylorT4}$ waveform model is used for the generating of the waveform. From the frequency profile, it is clear that the early inspiral part can be sampled with $128$ $\Hz$, whereas for the merger and ringdown parts, we can use a higher sampling frequency such as $2048$ $\Hz$ and $512$ $\Hz$. }
\label{Fig:Time-Vs-Freq-Evol-NSBH}
}
\end{figure} 

\section{Optimization of the template bank division}
\label{Subsec:Opt_Temp_div}
From the sub-section \ref{Subsec:Cost_SVD_Filter}, it is clear that for the template matrix factorization ($\SVD$, $\RMF$) based matched filtering scheme, one has to make a balance between the cost of filtering and the cost of reconstruction of the $\SNR$ time-series. There is always a trade-off between the filtering cost and reconstruction cost depending on the size of a template bank. For a large template matrix, the reconstruction cost is the dominant cost. Hence, the current online search pipeline dividing the whole template bank into sub-banks to optimally reduce the reconstruction cost. This trade-off entirely depending on the size of the bank, the number of sub-banks. Also, it depends on the relation between the number of basis vectors obtained from a full bank factorization and the total number of basis vectors obtained from all sub-banks factorization for a fixed $\avgSNRLoss$. It is expected that the number of basis vectors is small for a full bank decomposition in comparison to the total number of basis vectors obtained after the decomposing of all sub-banks. Therefore, in that case, the filtering cost is high if one has used the sub-banks division setup. Hence, for a large bank, only considering reducing reconstruction cost by prescribed the division of the bank into sub-banks may not always be optimal. Hence, it is crucial to analyze the total time complexity cost theoretically, including the filtering cost and the reconstruction cost for both the cases of the full bank and the sub-banks. 

Let a full bank contains $N_{T}$ number of templates, and each template has $N_{s}$ number of sample points. Also, let the whole bank can be divided into a $p$ number of sub-banks, and each sub-bank contains ${N_T}_{i}$ of templates. For simplicity, we have assumed that each sub-banks includes an equal number of templates ($b$). Hence $N_{T} = p\,{N_{T}}_{b}$, Let $k_{i}$ represent the number of required basis vectors for the $i^{\th}$ sub-bank corresponding to a fixed $\avgSNRLoss$. Reconstruction cost for each of the sub-bank is $k_{i} \, {N_T}_{b} \, N_{s}$. Filtering cost is $k_{i} \, N_{s} \, \log N_{s}$. Therefore the total cost for each of the sub-bank is $\big({N_T}_{b}\,k_{i}\,N_{s} + k_{i}\,N_{s}\,\log N_{s}\big)$. Therefore the total cost considering all the sub-banks is
\begin{equation}
\sum_{i = 1}^{p}\{{N_T}_{b} \, k_{i} \, N_{s} + k_{i} \, N_{s} \, \log N_{s} \} \, 
\label{Eq:TC_sub}
\end{equation}
Whereas, for a full-bank the reconstruction cost is $N_{T} \,k \, N_{s}$. Where $k$ is the number of the required basis for the whole bank corresponding to the same $\avgSNRLoss$ considered for each sub-banks. Therefore, The filtering cost is $k\, N_{s} \, \log N_{s}$ and the total cost for a full bank is 
\begin{equation}
N_{T} \,k \, N_{s} + k \, N_{s} \, \log N_{s}
\label{Eq:TC_full}
\end{equation}
Let $\gamma$ is the ratio between the total time complexity of matched filtering using full bank and sub-banks. Hence $\gamma$ is
\begin{equation}
\gamma  =  \frac{\sum_{i = 1}^{p}\{{N_{T}}_{b} \, k_{i} \, N_{s} + k_{i} \, N_{s} \, \log N_{s} \}}{N_{T} \,k \,N_{s} + k\,N_{s}\, \log N_{s}}
\label{Eq:TC_ratio}
\end{equation}
For simplicity, assume that all the sub-banks have the same number of basis vectors corresponding to a fixed $\avgSNRLoss$. Therefore the cost for every sub-bank is also the same. Thus The above Eq.\ref{Eq:TC_sub} can be written as follows:
\begin{equation}
p\Big({N_{T}}_{b} \, k_{i} \, N_{s} + k_{i} \, N_{s} \, \log N_{s} \Big)
\label{Eq:TC_sub1}
\end{equation}
The total number of basis considering all the sub-banks is $p \, k_{i}$. Generally $k \ll p \, k_{i}$. Therefore, we can consider $k$ as a $\beta$ fraction of $pk_{i}$ \text{i.e.} $k = \frac{pk_{i}}{\beta}$. Combining Eq.\ref{Eq:TC_full} $\&$ \ref{Eq:TC_sub1} and incorporating $k = \frac{pk_i}{\beta}$, one can redefine the ratio $\gamma$ as follows:
\begin{eqnarray}
\gamma & = & \frac{p\Big({N_{T}}_{b}\,k_{i}\,N_{s} + k_{i}\,N_{s}\,\log N_{s}\Big)}{N_{T}\,k\,N_{s} + k\,N_{s}\,\log N_{s}} \nonumber \\
& = & \frac{\beta \,\Big(N_{T} + p \,\log N_{s}\Big)}{p\,\Big(N_{T} + \,\log N_{s} \Big)} 
\label{Eq:TC_ratio1}
\end{eqnarray}
We can formulate three different cases based on the relation between $p$ and $\beta$. 
\begin{enumerate}
\item Case-{\text{I}}: Let us suppose if $p = \beta$, then For a large $N_{T}$ and small value of $p$, $\gamma \rightarrow 1$. This implies that the total cost for a sub-bank division is small compared to the full bank for this scenario. Whereas if $p$ is as large as $N_{T}$ then $\gamma \rightarrow (1 + \log N_{s})$. This shows that if each sub-bank contains a minimal number of templates. But this is not a feasible case as if $p$ becomes large, then $\beta$ has to be large as the underline assumption is $p = \beta$. Hence, for this case, either $p$ is small or large, the total cost for the sub-bank division-based matched filtering scheme is optimal. 
\item Case-{\text{II}}: Let $p < \beta : \beta = m \, p$, then from Eq.\ref{Eq:TC_ratio1}, 
\begin{equation}
\label{TC_caseII}
\gamma = \frac{m \big(N_{T} + p\,\log N_{s} \big)}{N_{T} + \log N_{s}}
\end{equation}
If $N_{T}$ is large, then $\gamma \rightarrow m$ no matter $p$ is small or large. Hence for this Case, the computational time for a full-bank decomposed-based matched filtering scheme is optimal.
\item Case {\text{III}}: Let $p > \beta : p = n \, q$, then from Eq.\ref{Eq:TC_ratio1}, 
\begin{equation}
\label{TC_caseIII}
\gamma = \frac{N_{T} + p\, \log N_{s}}{n \big(N_{T} + \log N_{s} \big)}
\end{equation}
for large $N_{T}$ and small value of $p$, $\gamma \rightarrow \frac{1}{n}$ whereas for $p = N_{T}$, $\gamma \rightarrow \frac{1+\log N_{s}}{n}$. For a substantial value of $p$, it seems to be observed that the computational cost for the full bank-based matched filtering scheme is less in comparison to the sub-banks-based filtering scheme. However, using the same argument described in the Case-{\text{I}}, it is clear that practically $p$ can't be the as same order as $N_{T}$. Hence, In this Case, the sub-bank division-based matched filtering scheme is optimal.   
\end{enumerate}
\begin{figure}[htbp]
\centering{
\includegraphics[width=.50\textwidth]{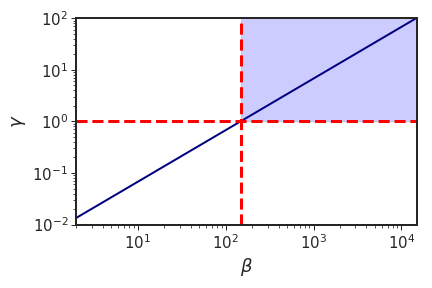}}
\caption{The figure shows the relation between the variables $\beta$ (defined in Ref. \cite{kulkarni2019random}, Figure-1) and $\gamma$. If $\beta = 1$, then it is implied that the number of required basis considering all the sub-banks (w.r.t a fixed $\avgSNRLoss$) is the same as the basis obtained from the full bank. The shaded region shows the possible configuration of $\beta$ and $\gamma$ in which the filtering cost will be less for the decomposition of the entire template bank than decompose into sub-banks and split banks together.}
\label{Fig:TC_ratio}
\end{figure}

After analyzing all possible relations between $p$ and $\beta$, It is clear that the total computational cost will be less in the case of the full bank in comparison to the sub-banks if and only if $p < \beta$ (Case-\text{II}). For a large template bank, one can conclude that the ratio $\gamma$ solely depends on the ratio between $\beta$ and $p$. Therefore, theoretically, it is easy to find those values of $\beta$ for which the computational cost for the full bank is less in comparison to the sub bank division. Still, practically, it is hard to directly find the relation between $p$ and $\beta$ without trials and testing the decomposition of the template matrix as a whole and into sub-banks. To provide a notion of this relation, we have consider an example in which, let us consider $N_{T} = 10^{5}$, $N_{s} = 10^{6}$ and $p = 200$ that means each sub-banks contain $500$ templates. in this set-up, $\gamma > 1$ only when $\beta > p ( = 200)$. Figure-\ref{Fig:TC_ratio} depicts the same relation. This relation implies that if the total number of basis vectors after decomposed the full bank is at least $200$ (= $p$) times less than the total number of basis vectors considering all the sub-banks, then the only total computational cost will be less for the full-bank factorized based matched filtering scheme. But for the other two cases, the reconstruction cost is the dominant cost. Therefore sub-bank division is the only known optimal solution. Thus, it is important to address the problem of reducing the reconstruction cost of the $\SNR$ time-series construction. For any matrix factorization-based matched filtering scheme, the set of bases are used to filter against the data, and the computation of the approximated $\SNR$ time-series depends on the multiplication of the coefficient vectors with the filter output. Therefore multiplication with the coefficient matrix is an essential step in this process and can not be ruled out. In General, the coefficient matrix obtained using $\SVD$ or $\RMF$ is a dense matrix that means the weights corresponding to the basis are non-zero. Thus the reconstruction cost is high. However, the $\MP$ algorithm allows us to obtain sparse coefficients for a set of bases. The use of the sparse coefficient matrix by replacing the dense one also incurred more approximation error in the waveform reconstruction and as well as in the $\SNR$ time-series construction. However, the approximation error is small. Therefore the sparse matrix can be used to construct the $\SNR$ time-series. In this way, we can reduce the reconstruction cost if we use a sparse coefficient matrix obtained from the $\MP$ algorithm in the place of a dense coefficient matrix using $\SVD$ or $\RMF$. 
\bibliographystyle{apsrev}
\bibliography{references}{}
\end{document}